\documentclass[a4paper]{spie}
\usepackage[]{graphicx}
\title{Phase Closure Image Reconstruction for Future VLTI Instrumentation}
\author{Mercedes E. Filho\supit{a}, Stephanie Renard\supit{b}, Paulo Garcia\supit{a,c}, Gilles Duvert\supit{d}, Gaspard Duchene\supit{e}, Eric Thiebaut\supit{b}, John Young\supit{f}, Olivier Absil\supit{g}, Jean-Phillipe Berger\supit{g}, Thomas Beckert\supit{h}, Sebastian Hoenig\supit{h}, Dieter Schertl\supit{h}, Gerd Weigelt\supit{h}, Leonardo Testi\supit{i}, Eric Tatuli\supit{i}, Virginie Borkowski\supit{j}, Micha\"el de Becker\supit{j}, Jean Surdej\supit{j}, Bernard Aringer\supit{k}, Joseph Hron\supit{k}, Thomas Lebzelter\supit{k}, Andrea  Chiavassa\supit{l}, Romano Corradi\supit{l}, Tim Harries\supit{m}
\skiplinehalf
\supit{a}Centro de Astrofisica da Universidade do Porto, Rua das Estrelas, 4150-762 Porto, Portugal \\
\supit{b}Observatoire de Lyon, 9 Av. Charles Andr\'e, 69561 Saint Genis Laval Cedex, France \\
\supit{c} Departamento de Engenharia F\'{\i}sica, Faculdade de
Engenharia, Universidade do Porto, Portugal \\
\supit{d}Laboratoire d'Astrophysique de Grenoble, Observatoire de Grenoble, BP 53, 38041 Grenoble Cedex 9, France \\
\supit{e}UC Berkeley, Astronomy Department, 601 Campbell Hall, Berkeley CA 94720-3411, USA \\
\supit{f}Cavendish Laboratory, Madingley Road, Cambridge CB3 OHE, UK\\
\supit{g}Universit\'e J. Fourier, CNRS, Laboratoire d'Astrophysique de Grenoble, UMR 5571, France;\\
\supit{h}Max-Planck Institute for Radioastronomy, Bonn, Germany;\\
\supit{i}INAF/Osservatorio di Astrofisica di Arcetri, Italy;\\
\supit{j}Institute of Astrophysics and Geophysics, Li\`ege, Belgium;\\
\supit{k}Institute of Astrophysics of the University of Wien, Austria;\\
\supit{l}Groupe  de  Recherche  en Astronomie  et Astrophysique du Languedoc,
Montpellier, France;\\
\supit{m}School of Physics, University of Exeter, UK;\\
}

\authorinfo{Further author information: (Send correspondence to M. E. F.)\\M. E. F.: E-mail: mfilho@astro.up.pt, Telephone: +351 6089 853}

\begin{document}  
 \maketitle 

\begin{abstract}

Classically, optical and near-infrared interferometry have relied on 
closure phase techniques to produce images. Such techniques allow us to 
achieve modest dynamic ranges.

In order to test the feasibility of next generation optical interferometers in the context of the VLTI-spectro-imager (VSI), we have embarked on a study of image reconstruction and analysis. Our main aim was to test the influence of the number of telescopes, observing nights and distribution of the visibility points on the quality of the reconstructed images. Our results show that observations using six Auxiliary Telescopes (ATs) during one complete night yield the best results in general and is critical in most science cases; the number of telescopes is the determining factor in the image reconstruction outcome.  

In terms of imaging capabilities, an optical, six telescope VLTI-type configuration and $\sim$200 meter baseline will achieve 4 mas spatial resolution, which is comparable to ALMA and almost 50 times better than JWST will achieve at 2.2 microns. Our results show that such an instrument will be capable of imaging, with unprecedented detail, a plethora of sources, ranging from complex stellar surfaces to microlensing events.


 \end{abstract}



\section{INTRODUCTION}
\label{sec:intro}  

Image reconstruction is a key problem in observational astronomy, in particular in optical interferometry, where it is needed to overcome existing ambiguities in the interpretation of simple visibility measurements.  


In an array of {\it N} telescopes, signals are combined in {\it $\frac{1}{2}$ $\times$ N $\times$ (N-1)} pairs or baselines to obtain  {\it $\frac{1}{2}$ $\times$ N $\times$ (N-1)} measurements called complex visibilities. These visibilities are related to the object brightness distribution via the van Cittert-Zernike theorem:

\begin{center}
$V(u,v) = \int \int I(x,y) \, exp[-2\,  \pi \, i \, (ux+vy)] dx \, dy$
\end{center}

\noindent where {\it x} and {\it y} are angular displacements on the plane of the sky with the phase center as origin, {\it I(x,y)} is the brightness distribution of the target and {\it u} and {\it v} are the position vectors of the baselines projected on a plane perpendicular to the source direction, which together define the {\it uv} plane. In practical terms, the better the sampling of the {\it uv} plane in terms of baseline length, position angle, and number of measurements, the more faithful the reconstructed image will be relative to the true brightness distribution. 

In order to recover the brightness distribution, several image reconstruction algorithms for interferometry have been developed: 
phase referencing using {\sc aips}, hybrid mapping using {\sc difmap} (Masoni et al. 2005, Masoni 2006), 
the building block bispectrum method (Hofmann \& Weigelt 1993), {\sc mira} (Thiebaut 2005), {\sc bsmem} (Baron 2007; private communication), 
among others.

Closure phase imaging is the standard method for imaging in optical interferometry. It has been applied since the 1980's, initially to speckle interferometry (Weigelt et al. 1998) and masked instrumentation and lately to optical interferometry with separate telescopes.
This technique relies on two quantities related to the complex visibility –- the squared visibility and the triple product -– to perform image reconstruction. 

The squared visibility or power-spectrum is defined as:

\begin{center}
$V(u_i,v_i) \, V^*(u_i,v_i)  = |V(u_i,v_i)|^2$
\end{center}

\noindent and the triple product or bi-spectrum:

\begin{center}
$T(u_i,v_i,u_j,v_j) = V(u_i,v_i) \, V(u_j,v_j) \, V^*(u_i+u_j,v_i+v_j)$
\end{center}

\noindent where {\it u$_i$+u$_j$=u$_k$, v$_i$+v$_j$=v$_k$, u$_i$, v$_j$, u$_i$ ,v$_j$}, are vectors defining the position of three baselines and the asterisk denotes the complex conjugate. The triple product is the product of the complex visibilities on the baselines forming a closed loop joining any three telescopes (Fig. 1). The modulus of the triple product is the triple product amplitude and the argument is the closure phase. The closure phase is independent of atmospheric-induced phase variations.

\begin{figure}[ht!]
   \begin{center}
   \begin{tabular}{c}
   \includegraphics[height=5cm]{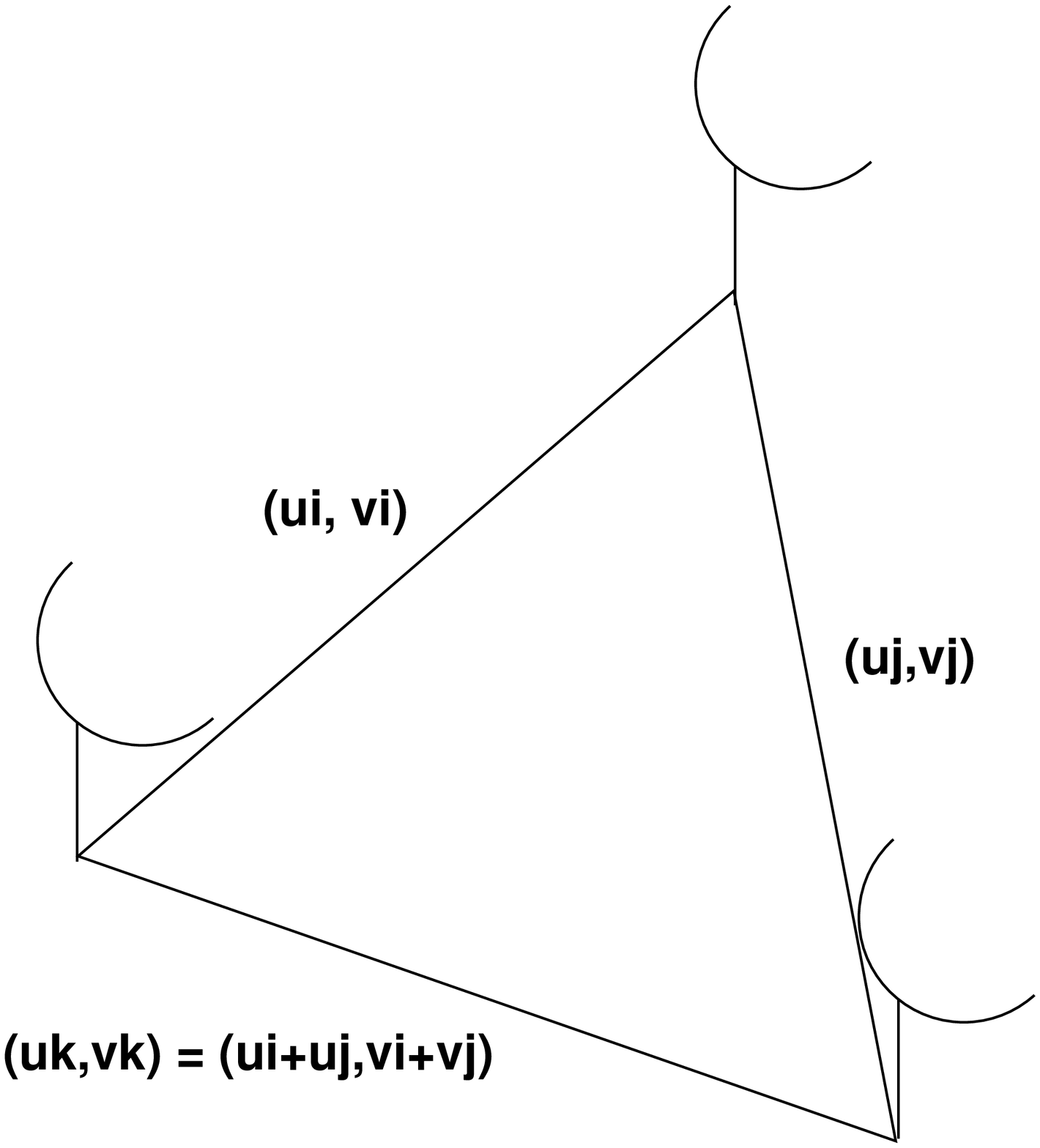}
   \end{tabular}
   \end{center}
   \caption[example] 
   { \label{fig:example} }
\end{figure}
    
\section {Array Configuration}

AMBER was the first VLTI instrument to use beam recombination for three telescopes. Next generation instruments like VSI (Malbet et al. 2006) and Matisse (Lopez et al. 2006) will use a larger number of baselines. In order to test the importance of the number, type and configuration of the telescopes and number of observing nights on the image reconstruction, we have adopted  three telescope setups using the VLTI as a template. The setups were chosen to optimize the {\it uv} plane coverage, while taking into consideration such limitations as the delay lines. 

The following configurations were adopted:

\begin{figure}[ht!]
   \begin{center}
   \begin{tabular}{c}
   \includegraphics[height=8cm,angle=-90]{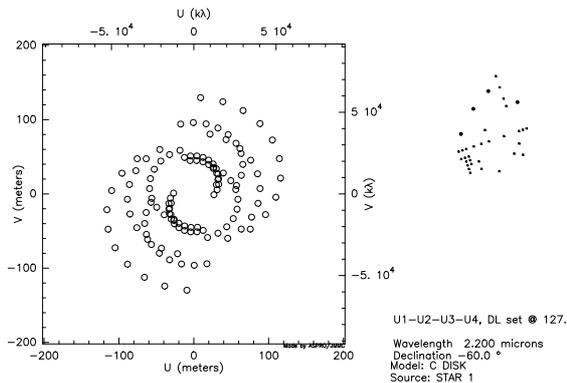}
   \end{tabular}
   \end{center}
   \caption[example] 
   { \label{fig:example} 4 UT $\times$ 1 night configuration.}
\end{figure}

\begin{figure}[h!]
   \begin{center}
   \begin{tabular}{c}
   \includegraphics[height=8cm,angle=-90]{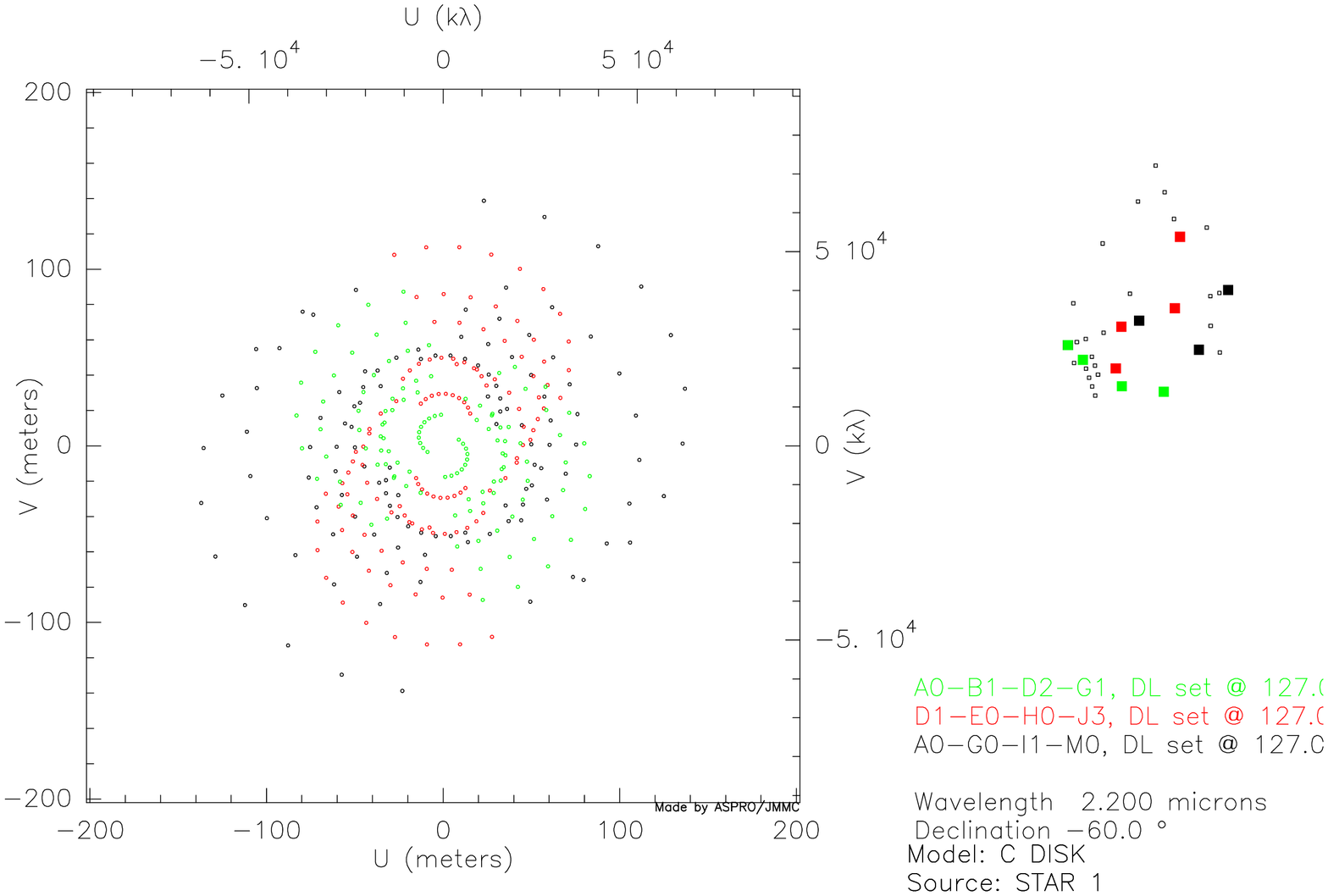}
\includegraphics[height=8.5cm,angle=-90]{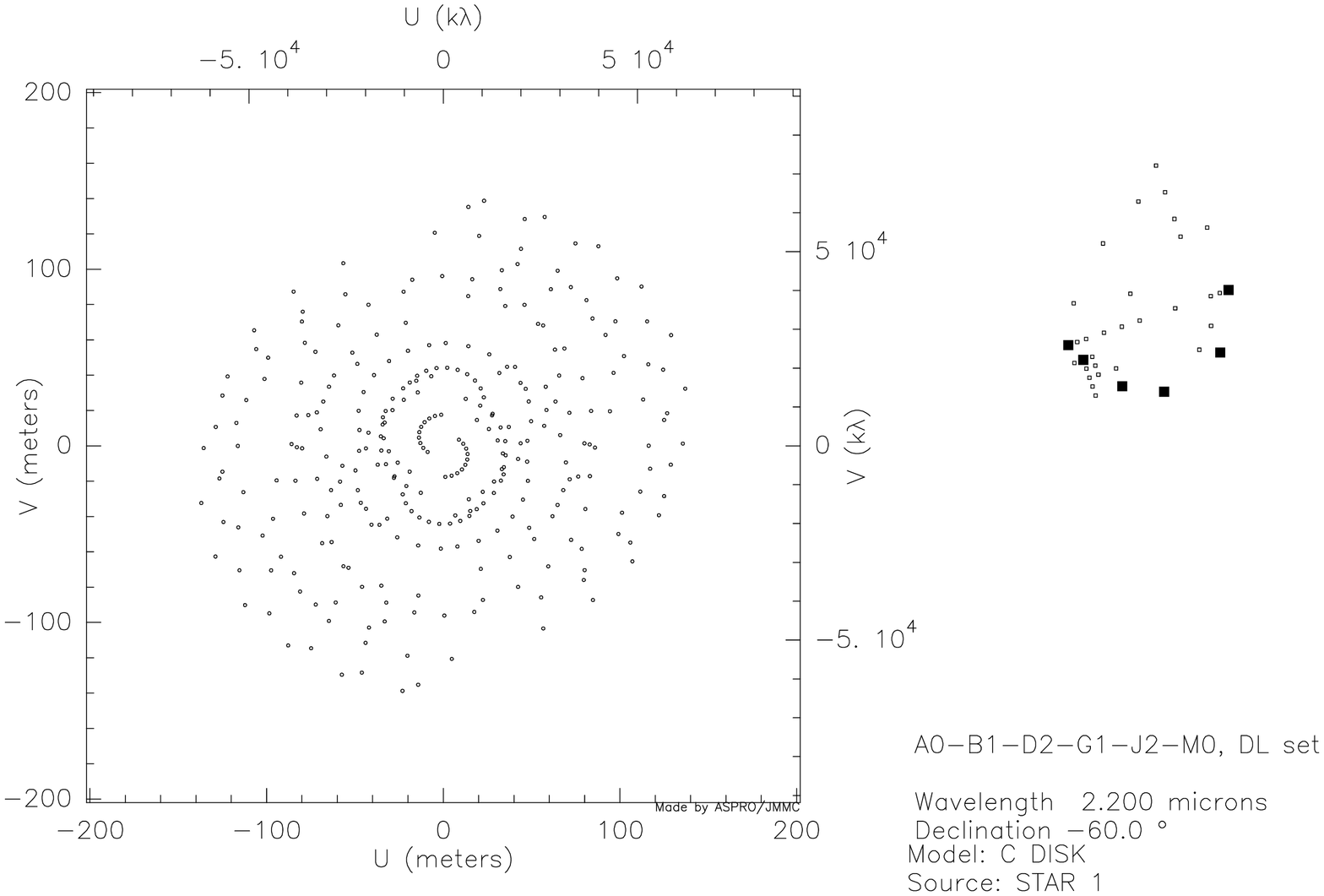}
   \end{tabular}
   \end{center}
   \caption[example] 
   { \label{fig:example} 4 AT $\times$ 3 night configuration (left). 6 AT $\times$ 1 night configuration (right).}
\end{figure}

\begin{itemize}

\item 4 UTs $\times$ 1 night (U1-U2-U3-U4) –- faint source configuration (Fig. 2);

\item 4 ATs $\times$ 3 nights (A0-B1-D2-G1, D1-E0-H0-J3, A0-G0-I1-M0) –- the minimum imaging $uv$ coverage consistent with the current VLTI capabilities (Fig. 3; left);

\item 6 ATs $\times$ 1 night (A0-B1-D2-G1-J2-M0) –- a 6 telescope extended configuration with one night observation; has less uv points than the 4 AT x 3 nights configuration (Fig. 3; right).

\end{itemize}

\section {Noise Model}

The noise model estimates the uncertainties on squared visibilities and closure phases assuming an instrument following a multi-axial recombination scheme with a fringe tracker (FT; Jocou et al.2007). 



The quantity we procure is the total number of detected photoevents per baseline in 5 minutes in the photometric channel:

\begin{center}
$N_p = \frac{N_{total} \times (1-f) \times 5 \times N_{obs}}{N_{base}}$
\end{center}

\noindent where $f$ is the fraction of the beam that gores into the interferometric channel (90\%), $N_{obs}$ is the number of observations per minute, $N_{base}$ is the number of baselines and $N_{total}$ is the total number of detected photevents per integration point in all channels:

\begin{center}
$N_{total} = F_0 \times 10^{-0.4 mag} \times t_{int} \times N_{tel} \times \pi \times R^2 \times \Delta \lambda \times trans \times Strehl$
\end{center}

\noindent Here $F_0$ is the photon flux of a zeroth magnitude star (Jocou et al. 2007), $mag$ is the object magnitude in the observing band, $t_{int}$ is the integration time (0.05 s), $N_{tel}$ is the number of telescopes (6 or 4), $R$ is the radius of the telescopes (4.1 for UTs and 0.9 for ATs) assumed for simplification to have no central hole, $\Delta \lambda$ is the spectral bandwidth (chosen), $trans$ is the total instrument transmission including quantum efficiency (Jocou et al. 2007), and $Strehl$ is the Strehl ratio and depends on wavelength (Jocou et al. 2007).

The correlated flux per baseline is given by:
 
\begin{center}
$F_{cor} = \frac {N_i \times V'}{2}$
\end{center}

\noindent where $V'$ is the instrumental visibility loss (80\%) and FT loss (90\%) corrected visibility and $N_{i}=N_{total} \times f \times 5 \times N_{obs} / N_{base}$, is the detected number of photoevents per baseline in 5 minutes in the interferometric channel. 

The signal-to-noise ratio of the squared correlated flux ($SNR$) is a complex function of $F_{cor}$, $N_i$, $V'$, $N_{tel}$, $N_{pix}$ the number of pixels needed to code the interferometric channel (600) and $\sigma$ the detector readout noise (15 e-). 

\noindent Therefore, the uncertainty on the squared visibility is:

\begin{center}
$error_{V^2} = V'^2 \times {\sqrt {SNR^{-2}+2 \times (N_p + N_{pix p} \, \sigma^2 / N_p^2)}} $
\end{center}

\noindent where $N_{pix p}$ is the number of pixels needed to code the photometric channel (4).

The photon noise on the closure phase ($Phot$) is a function of $N_{tel}$, $N_i$, $V_{123}$, $V_1$, $V_2$, $V_3$, $V_{12}$, $V_{23}$, $V_{13}$, where the $V$ terms correspond to visibilities in a combination of one, two or three telescopes. The detector noise on the closure phase ($Det$) depends on $N_{tel}$, $N_i$, $N_{pix}$, $\sigma$ and the $V$ terms.

Therefore, the total noise on the closure phase is estimated as:

\begin{center}
$error_{T} = \sqrt {\frac{Phot+Det}{5 \times N_{obs}}}$
\end{center}


\section {OIFITS File Generation}

The advent of optical interferometry has lead to the need for a common data format based on the Flexible Image Transport System (FITS) in order to facilitate data reduction and the combination of various types of interferometer data. The result was the creation of the OI Exchange Format or OIFITS format (Pauls et al. 2004). Interferometer projects supporting this standard include COAST, NPOL, IOTA, VLTI, PTI and the Keck Interferometer. 

OIFITS format is designed so that different categories of information are stored in distinct “tables” within a file and can be cross-referenced one to another. Each OIFITS file “table” stores specific parameters that include interferometric observables –- complex visibility (OI-VIS), squared visibilities (OI-VIS2), and the triple product (bispectrum/closure phase; OI-T3) –- and ancillary data like Universal Time measurements, spectral bandpass and wavelengths (OI-WAVELENGTH), target identification (OI-TARGET) and telescope coordinates (OI-ARRAY).

Synthetic images were provided by the science case groups (Garcia et al. 2007) and imported into {\sc aspro}, an image simulation tool originally created for IRAM.  Assuming the objects were located at a $-$60 degree declination angle, typical object magnitudes were used to generate noisy, squared visibilities and closure phases.
We assumed an one hour per calibrated measurement. The actual on-source integration time is, however, 10-15 minutes per hour due to overheads. The total integration time assumes an entire transit (9 hours).

\section {Phase Closure Theory}

\begin{figure}[h!]
   \begin{center}
   \begin{tabular}{c}
   \includegraphics[height=3.5cm]{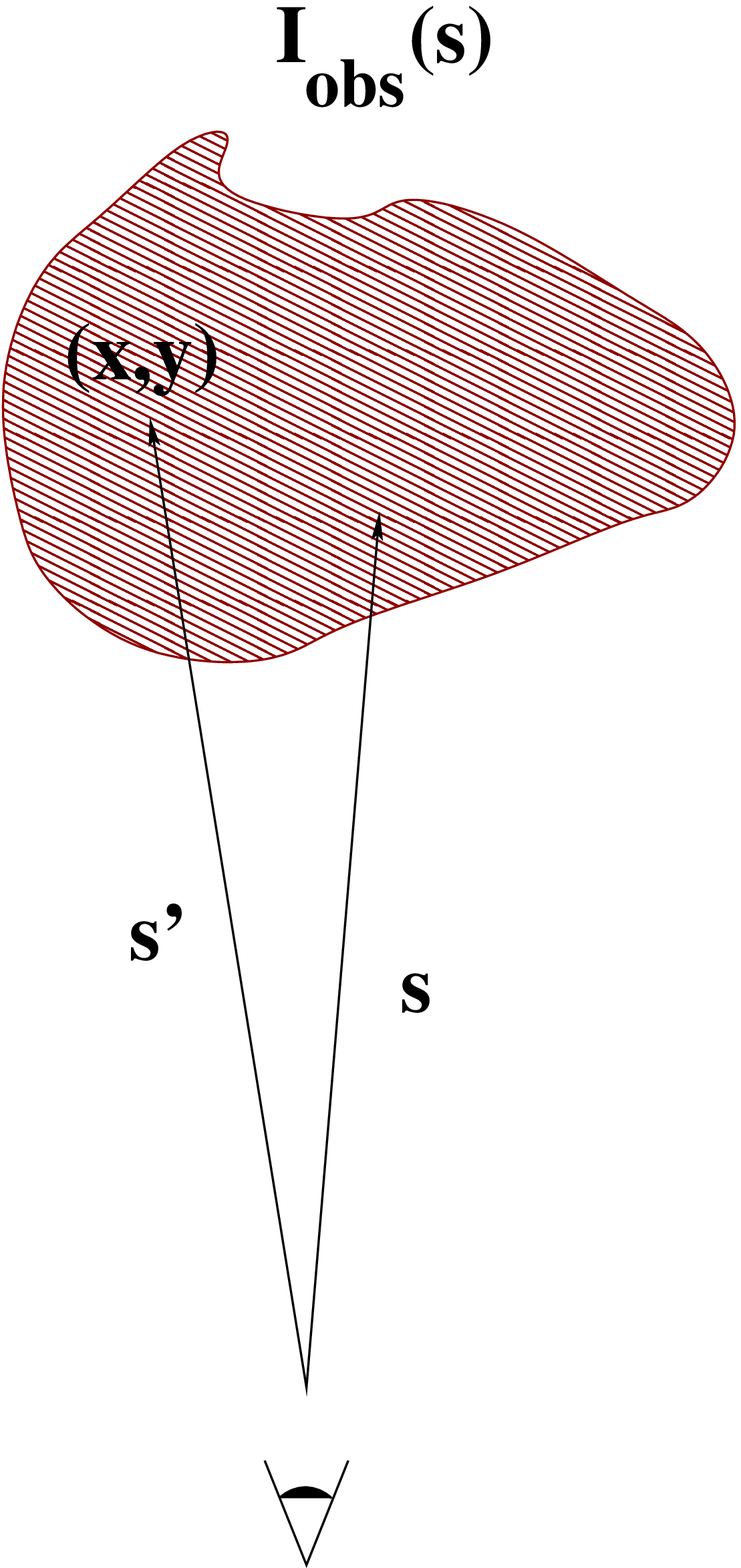}
   \end{tabular}
   \end{center}
   \caption[example] 
   { \label{fig:example} }
   \end{figure} 

Given an incoherent source, the observed brightness distribution in a direction $\vec{s'}$ is given by (Fig. 4):

\begin{center}
$I^{obs}(\vec{s'}) = I^{obs}(x,y) = PSF(x,y) * I^{true}(x,y) + N(x,y)$
\end{center}

\noindent where {\it PSF(x,y)} is the point spread function, {\it I$^{true}$(x,y)} is the true object brightness distribution, {\it N(x,y)} is the noise and the asterisk denotes convolution.

In practice, interferometry does not make measurements in the image plane but in Fourier space. The relevant quantity is called the complex visibility and is measured at each {\it uv} point, the position vector of the baseline on a plane perpendicular to the source direction:

\begin{center}
$V^{obs}(u,v) = V^{true}(u,v) \times S(u,v) + N'(u,v)$
\end{center}

\noindent where {\it S(u,v)}, the sampling function, is the Fourier transform of the {\it PSF(x,y)}, {\it V$^{true}$(x,y)}, the true visibility, is the Fourier transform of the true brightness distribution {\it I$^{true}$(x,y)} and {\it N'(u,v)} is the noise in the Fourier space.

The van Cittert-Zernike theorem states that the true brightness distribution can be obtained by the inverse Fourier transform and deconvolution of the observables:

\begin{center}
$I^{true}(x,y) * PSF(x,y) = \int \int V^{true}(u.v) \times S(u,v) \, exp \, [2 \pi i (ux+vy)]\, du \, dv$
\end{center}

The role of image reconstruction is to obtain the best approximation, {\it I$^{aprox}$(x,y) $\sim$ I$^{true}$(x,y)}, to the true brightness distribution. For phase closure image reconstruction we want to find a solution which has the maximum probability given the data. The solution {\it I$^{aprox}$(x,y)} involves minimizing the value of a penalty term subject to some condition:

\begin{center}
$P(z) = P_L(z) + P_{prior}(z) = P_L(z) + \mu R(z)$
\end{center}

\noindent where {\it z} represents the intensity of the pixels, {\it P$_L$} is the likelihood penalty, {\it $P_{prior}$} is the prior penalty, {\it R(z)} is the regularization term and $\mu$ is a multiplier (hyperparameter) tuned so that at the solution, the likelihood terms are equal to their expected values.  {\it P$_L$} enforces agreement with the data, while {\it P$_{prior}$} provides information where the data fail to do so, in particular in regions where the noise dominates or where the data are missing. {\it P$_{prior}$} is responsible for the so-called regularization of the inverse problem. 

The solution is therefore found by minimizing the likelihood penalty term for squared visibilities and triple products under a prior constraint:

\begin{center}
$P_L = P_{V^2} P_{T} \propto -\exp{\frac{\chi^2_{V^2}+\chi^2_{T}}{2}}$
\end{center}

\noindent where {\it $\chi^2_{V^2}$} is the likelihood term with respect  to the squared visibility data and {\it $\chi^2_T$} to the triple product. Typically, the data penalties are defined assuming the measurements follow Gaussian statistics:
				
\begin{center}
\[
  \chi^2_{V^2}(z) = \sum \frac{1}{\sigma^2_{V^2}} \, \left(
    V^2_{\mathrm{data}} - V^2_{\mathrm{model}}
  \right)^2
\]

\[
  \chi^2_{\mathrm{T}}(z) = \sum \frac{1}{\sigma^2_{\mathrm{T}}}
  \, \left\vert
    \mathrm{e}^{\mathrm{i}\,\phi^{\mathrm{data}}_{\mathrm{T}}} -
    \mathrm{e}^{\mathrm{i}\,(\varphi_1 + \varphi_2 - \varphi_3)}
  \right\vert^2
\]

\end{center}

\noindent where {\it V$^2$} refers to the squared visibilities (data and model), {\it $\phi_{T}^{data}$} to the measured closure phase, {\it $\varphi_1$, $\varphi_2$,  $\varphi_3$} to three individual phases involved in the phase closure and {\it $\sigma$} to the respective standard deviations.

Regularization can enforce agreement with some preferred and/or exact properties of the solution. Because we want to favor compact sources, smoothness is a very common regularization constraint:

\begin{center}
$R(z) = \sum \, w \, |z|^2$ 
\end{center}

\noindent where the regularization weights {\it w} are chosen in order to ensure spectral smoothing, that is, it enforces smoothness of the Fourier spectrum of the image and enforces compactness of the brightness distribution of the field of view. 








We have implemented the Multi-Aperture Reconstruction Algorithm ({\sc mira}; version 0.7; April 2008) as developed by Eric Thiebaut with support from the Jean-Marie Mariotti Center (JMMC). {\sc mira} works in the {\sc yorick} and C platform and is optimized to handle optical interferometric data with sparse {\it uv} coverage (Thiebaut 2005). Typically, {\sc mira} implements several types of regularization of which smoothness and maximum entropy are some examples. 




\section {Image Analysis}




       






Because {\sc mira} typically superresolves in the image reconstruction, we have compared our reconstructed images with the synthetic images provided by the science groups. Relative astrometry information was obtained for the images using {\sc ds9}. Photometry of the image components was performed using {\sc iraf} procedure {\sc phot}. $SNR_{V^2}$ and $SNR_{T}$ correspond to the range of signal-to-noise ratios in the squared visbilities and closure phases measured in the synthetic images.










\section {Closure Phase Image Reconstruction}

\begin{figure}[h!]
   \begin{center}
   \begin{tabular}{c}
   \includegraphics[height=4cm]{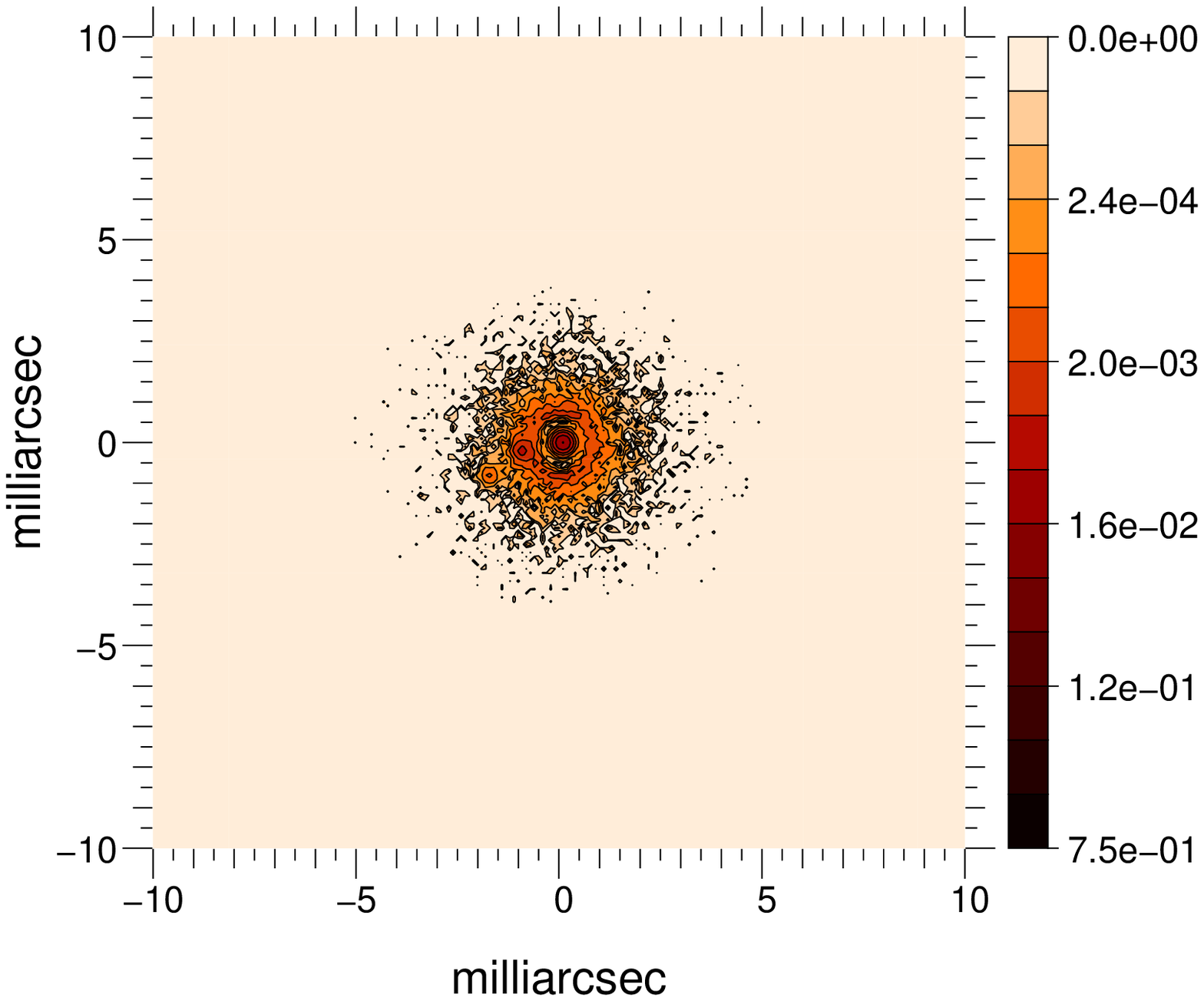}
  \includegraphics[height=4cm]{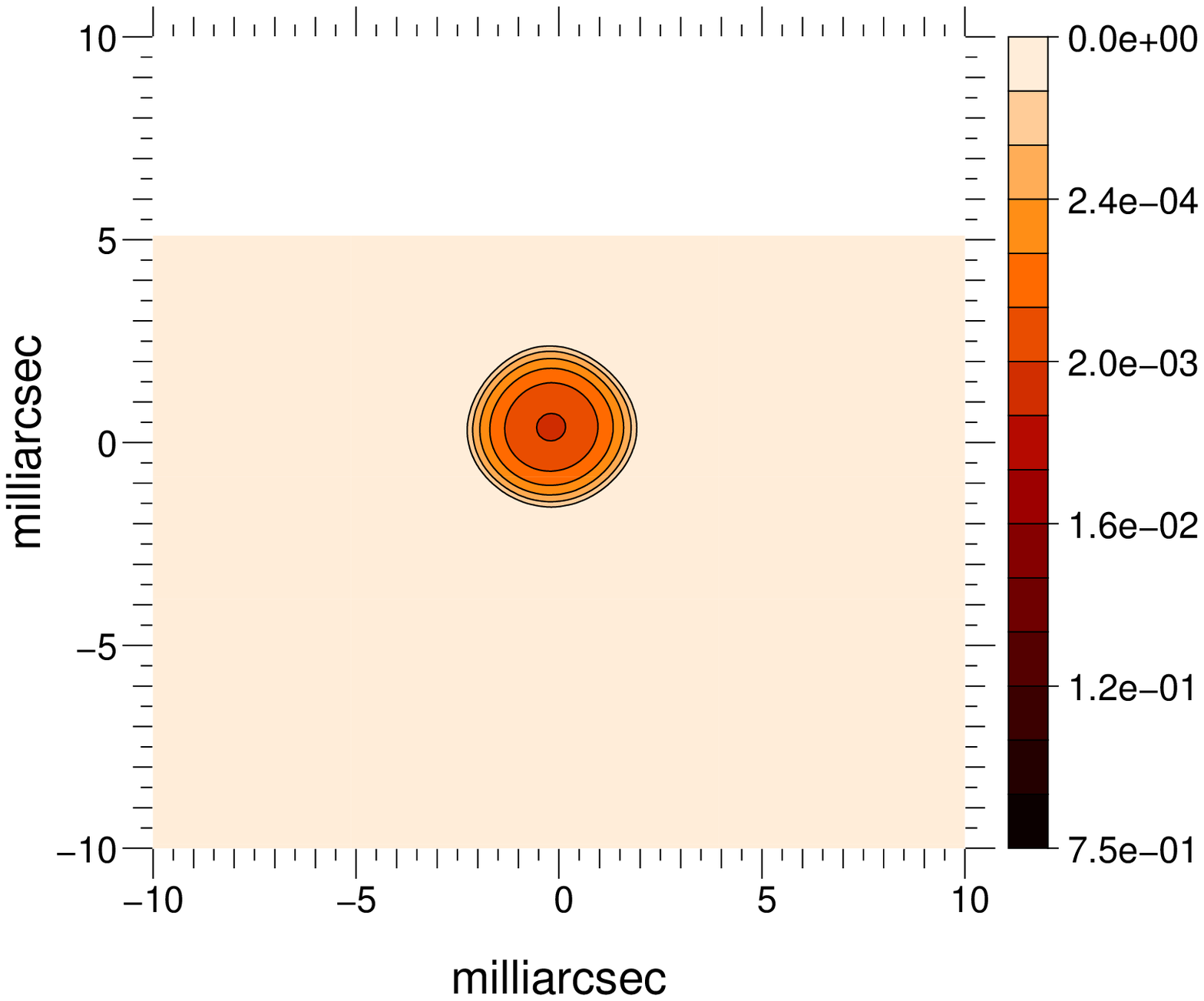}
   \end{tabular}
   \end{center}
   \caption[AGN] 
   { \label{fig:example} A simulated AGN with torus, 0.1 mas/pixel sampling; {\sc mira} reconstruction 4 UT $\times$ 1 night configuration, smoothness
regularization, 0.1 mas/pixel sampling.}
\end{figure}

\begin{table}[h!]
\begin{center}
\footnotesize
\caption {AGN. Diameter units are in pixels.}
\begin{minipage}[c]{63mm}
\begin{tabular} {l | c c c c}

\hline

    & Image   &         {\sc mira} 4 UT 	\\
\hline

flux nucleus	 & 22.3\%	&     -         \\
		
flux inner diameter  & 55.0\%  &   -	      \\

flux outer diameter &  17.6\% &   91.6\%\\ 

flux torus & 5.1\% & 8.4\% \\

\hline 

nuclear diameter  & 7  &   - \\

inner diameter & 30 & - \\

outer diameter & 75 & 70 \\

torus diameter & 260 & - \\

\hline

SNR$_{V^2}$    &  -   & 553-704 \\

SNR$_{T}$      & -    & 0.3-0.4 \\


\hline

\end{tabular}
\end{minipage}
\end{center}
\end{table}

















\begin{figure}[h!]
   \begin{center}
   \begin{tabular}{c}
\includegraphics[height=4cm]{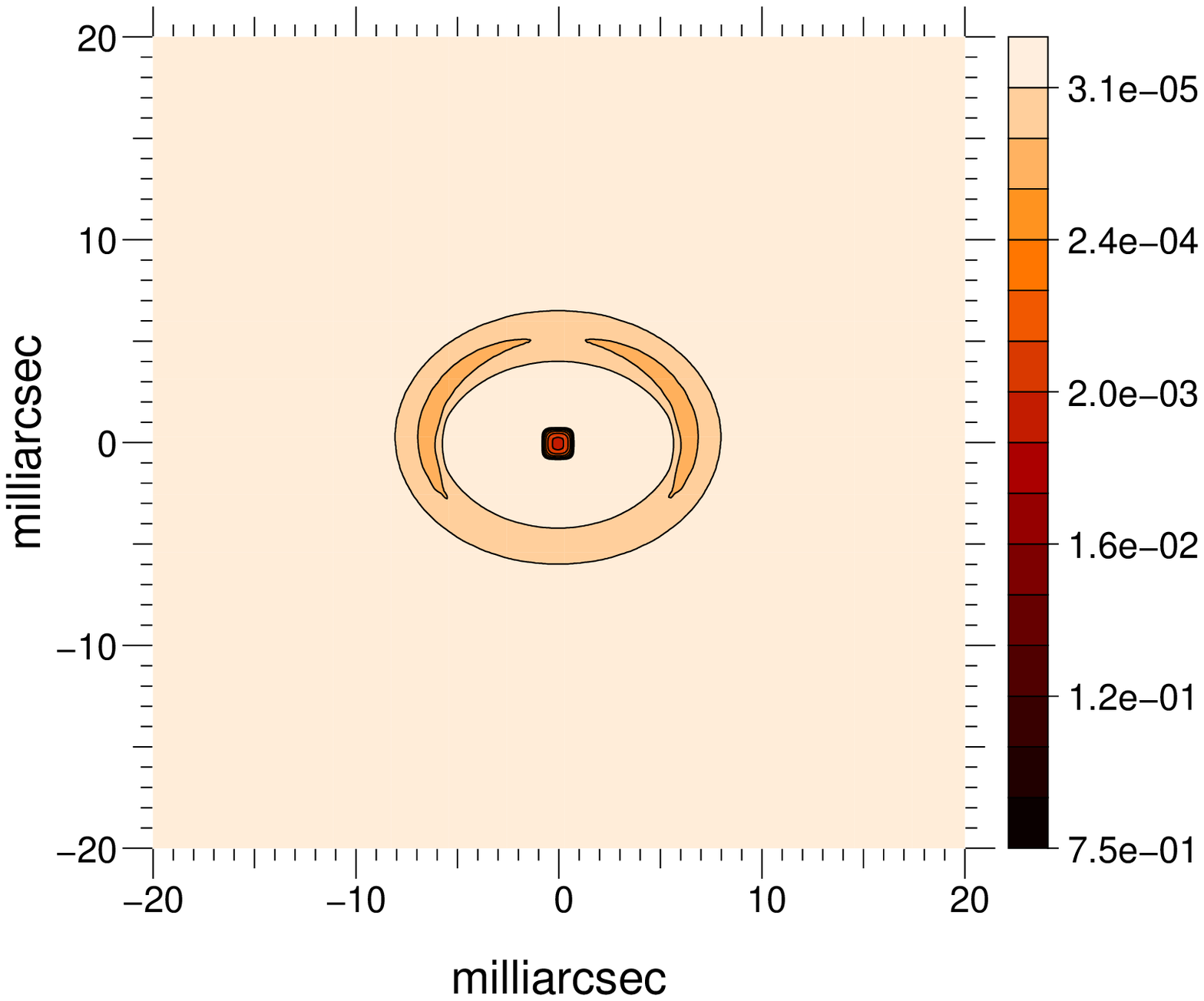}
\includegraphics[height=4cm]{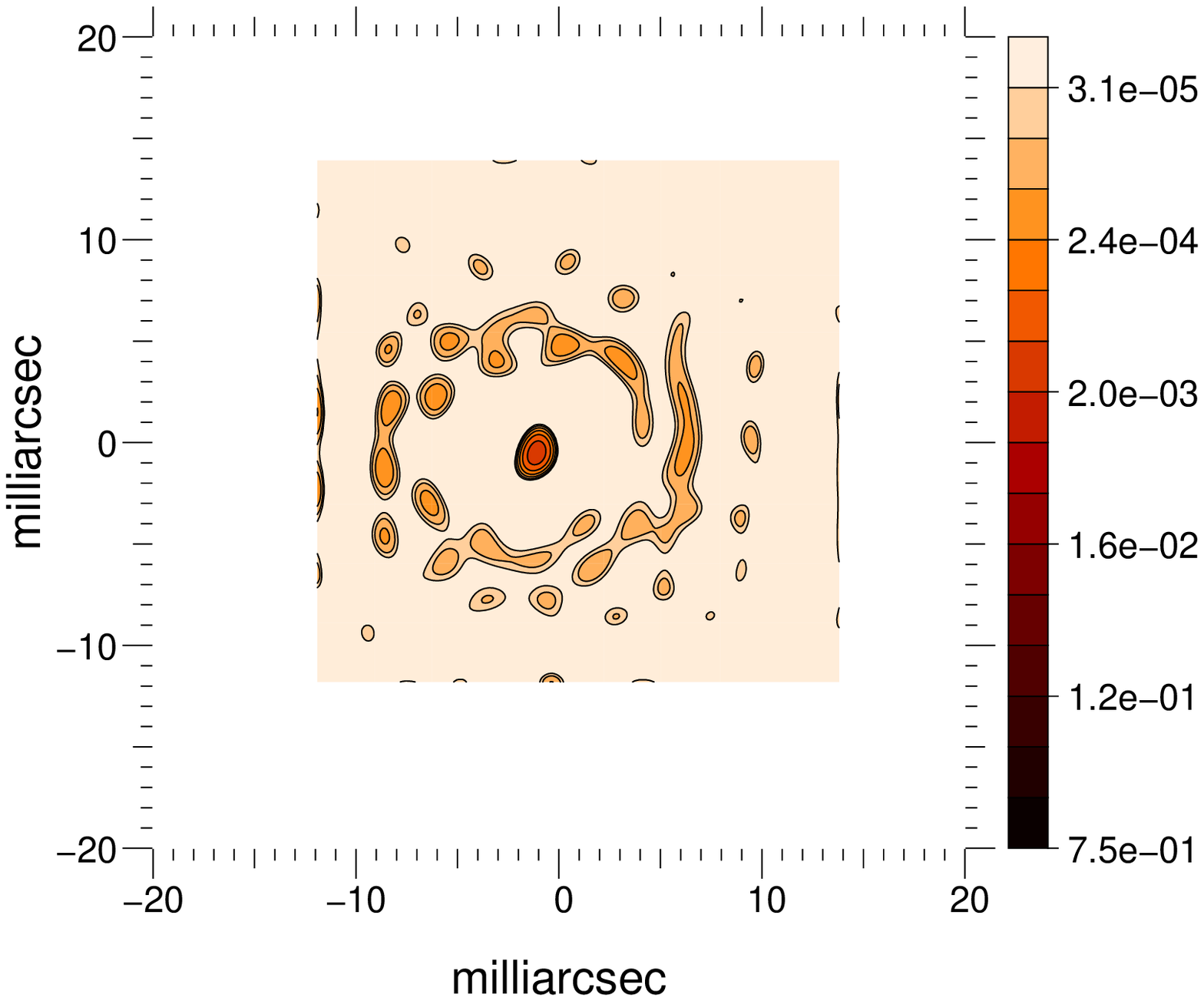}
\includegraphics[height=4cm]{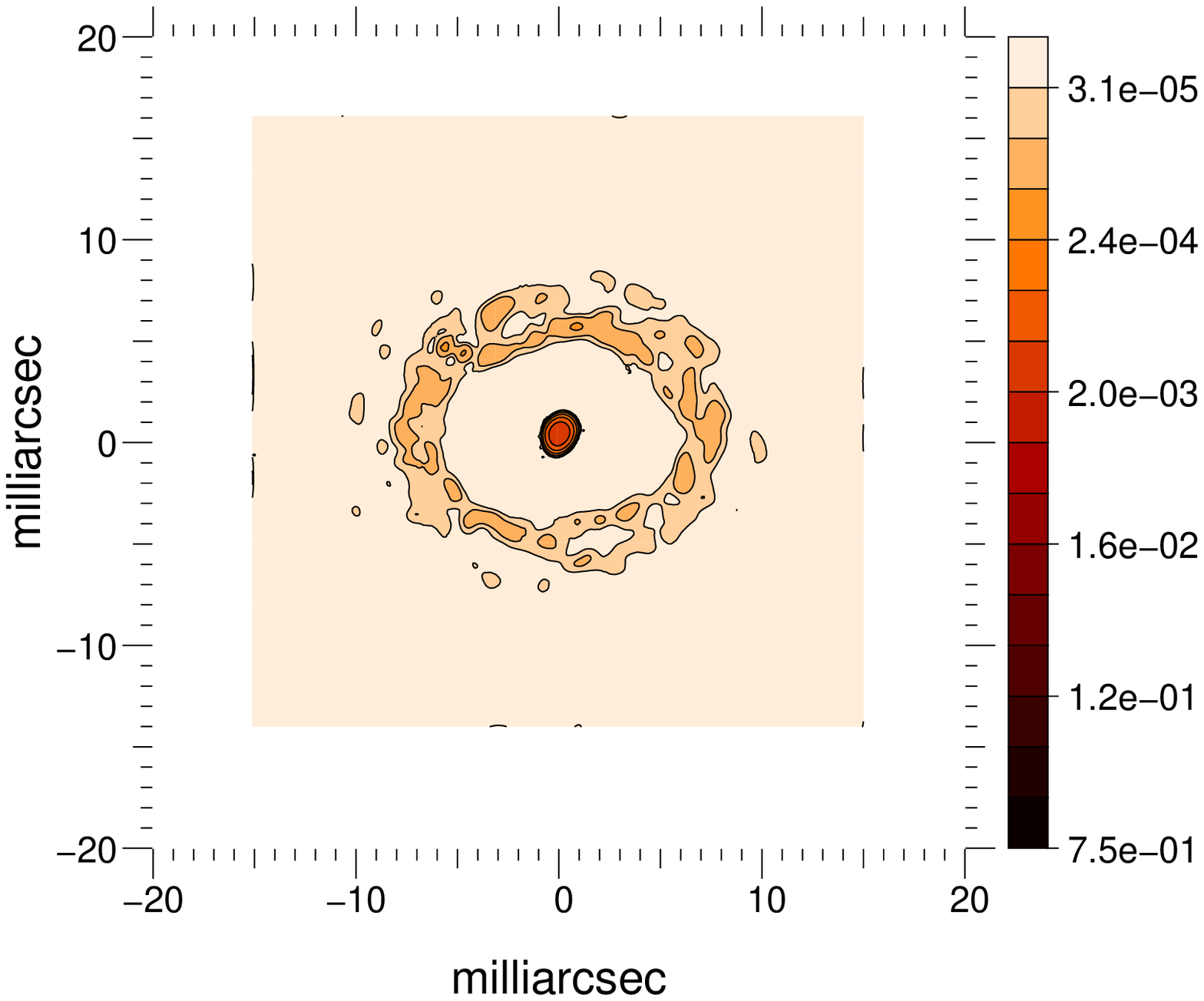}
\end{tabular}
   \end{center} 
   \caption[Evolved Star] 
   { \label{fig:example} A low mass evolved star system with an outflow, 0.1 mas/pixel sampling; {\sc mira} reconstruction 4 AT $\times$ 3 nights configuration, smoothness regularization, 0.1 mas/pixel sampling; {\sc mira} reconstruction 6 AT $\times$ 1 night configuration, smoothness regularization, 0.1 mas/pixel sampling.}
\end{figure}

\begin{table}[h!]
\begin{center}
\footnotesize
\caption{Evolved Star. Diameter units are in pixels.}
\begin{minipage}[c]{90mm}
\begin{tabular} {l | c c c }
\hline

      &	Image                   &         {\sc mira} 4 AT 3  &     {\sc mira} 6 AT  \\

\hline

flux star	&    25.1\%                 &   22.3\%       &	22.5\%      	\\

flux wind	&     74.9\%               &	77.7\%       	&  77.5\%          \\

ratio star/wind	&       0.3                     &  0.3         &   0.3 \\

\hline 

star diameter & 10  & 15 & 12 \\

inner wind diameter &  60 $\times$ 40 & 50 $\times$ 30 & 60 $\times$ 40  \\

outer wind diameter &  100 $\times$ 80 & 90 $\times$ 80 & 100 $\times$ 80 \\

\hline

SNR$_{V^2}$         &      -           & 63-2336      & 38-1706 \\

SNR$_{T}$           &      -           & 0.01-0.2     & 1.0-44 \\  


\hline
  
\end{tabular}  
\end{minipage}
\end{center}
\end{table}

\begin{figure}[h!]
   \begin{center}
   \begin{tabular}{c}
\includegraphics[height=4cm]{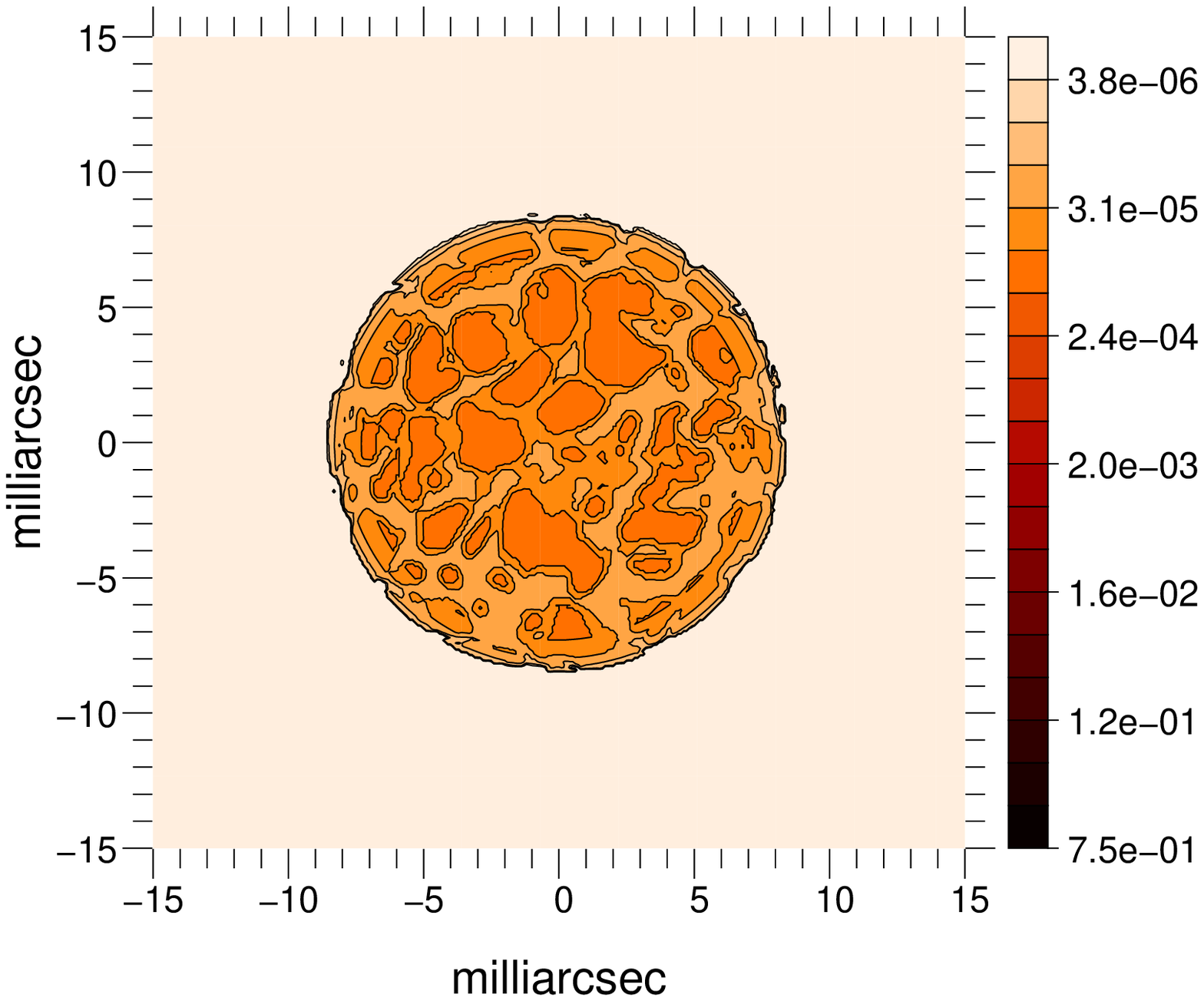}
\includegraphics[height=4cm]{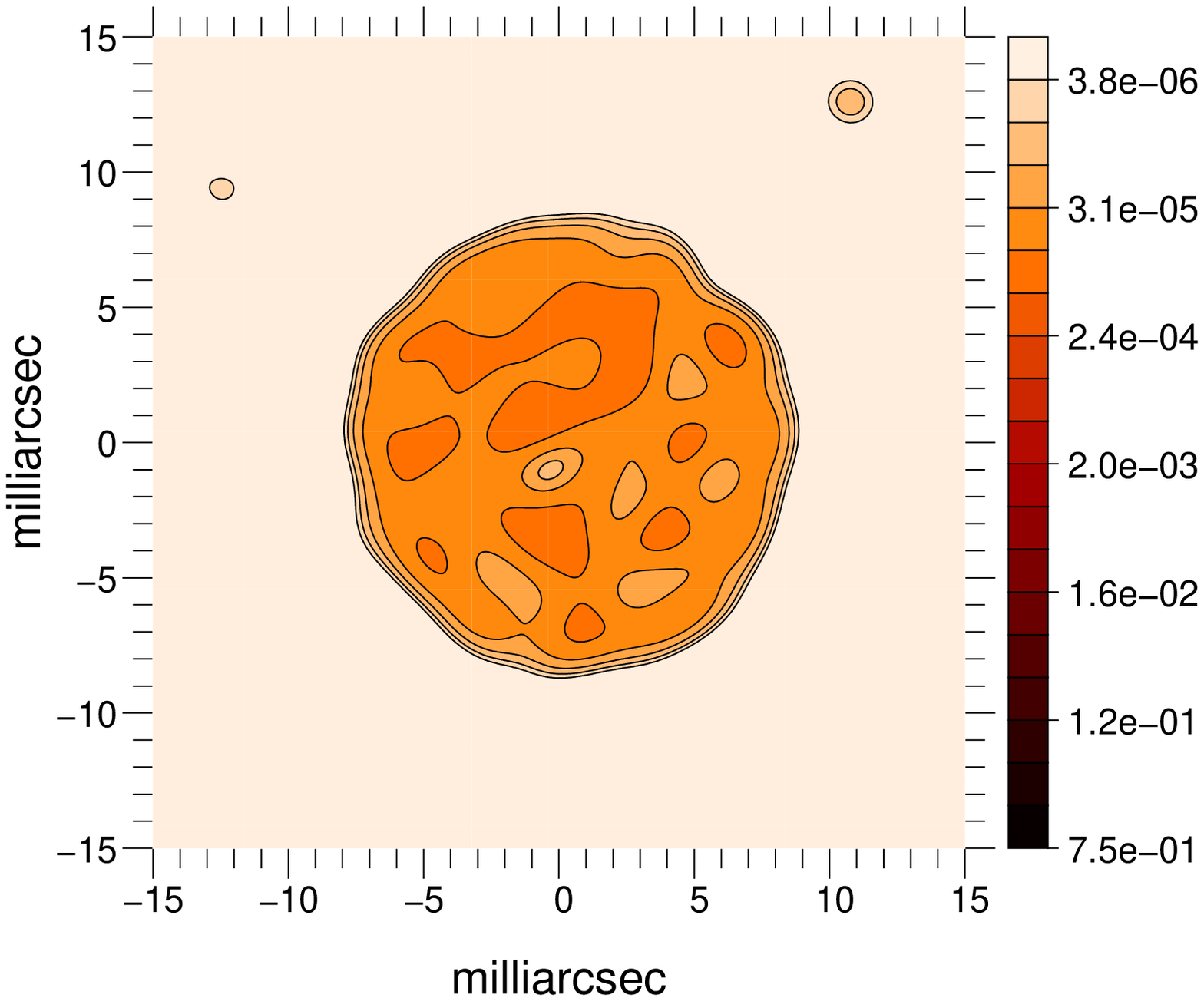}
\includegraphics[height=4cm]{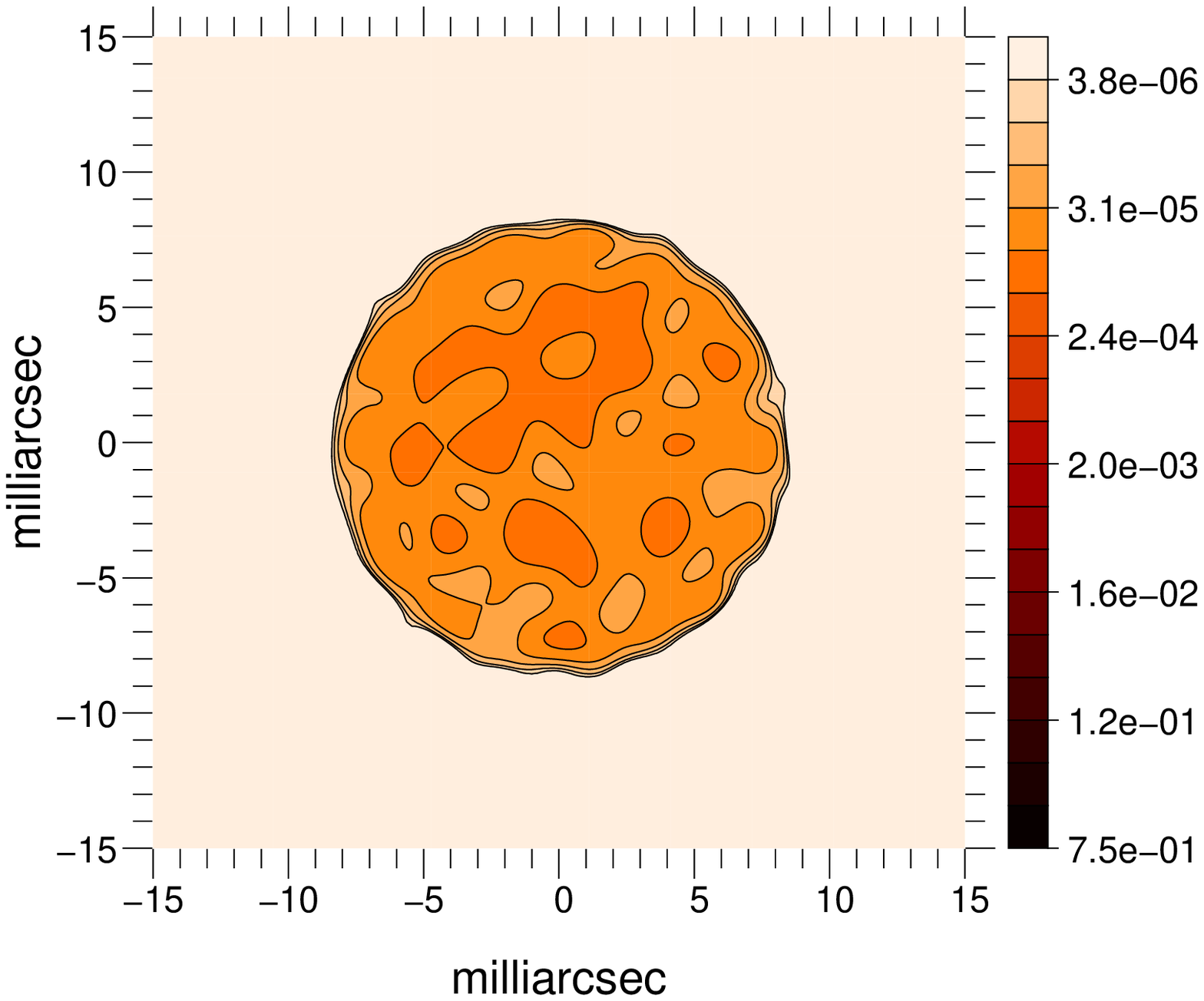}
   \end{tabular}
   \end{center}
   \caption[SG_surface] 
 {\label{fig:example} A simulated stellar surface of a M giant, 0.1 mas/pixel sampling; {\sc mira} reconstruction 4 AT $\times$ 3 nights configuration, smoothness regularization, 0.1 mas/pixel sampling, SNR$_{v^2}$=1.5-541, SNR$_{T}$=0.01-05; {\sc mira} reconstruction 6 AT $\times$ 1 night configuration, smoothness regularization, 0.1 mas/pixel sampling, SNR$_{V^2}$=0.1-600, SNR$_{T}$=0.1-115.}
\end{figure}




 





\begin{figure}[h!]
   \begin{center}
   \begin{tabular}{c}
\includegraphics[height=4cm]{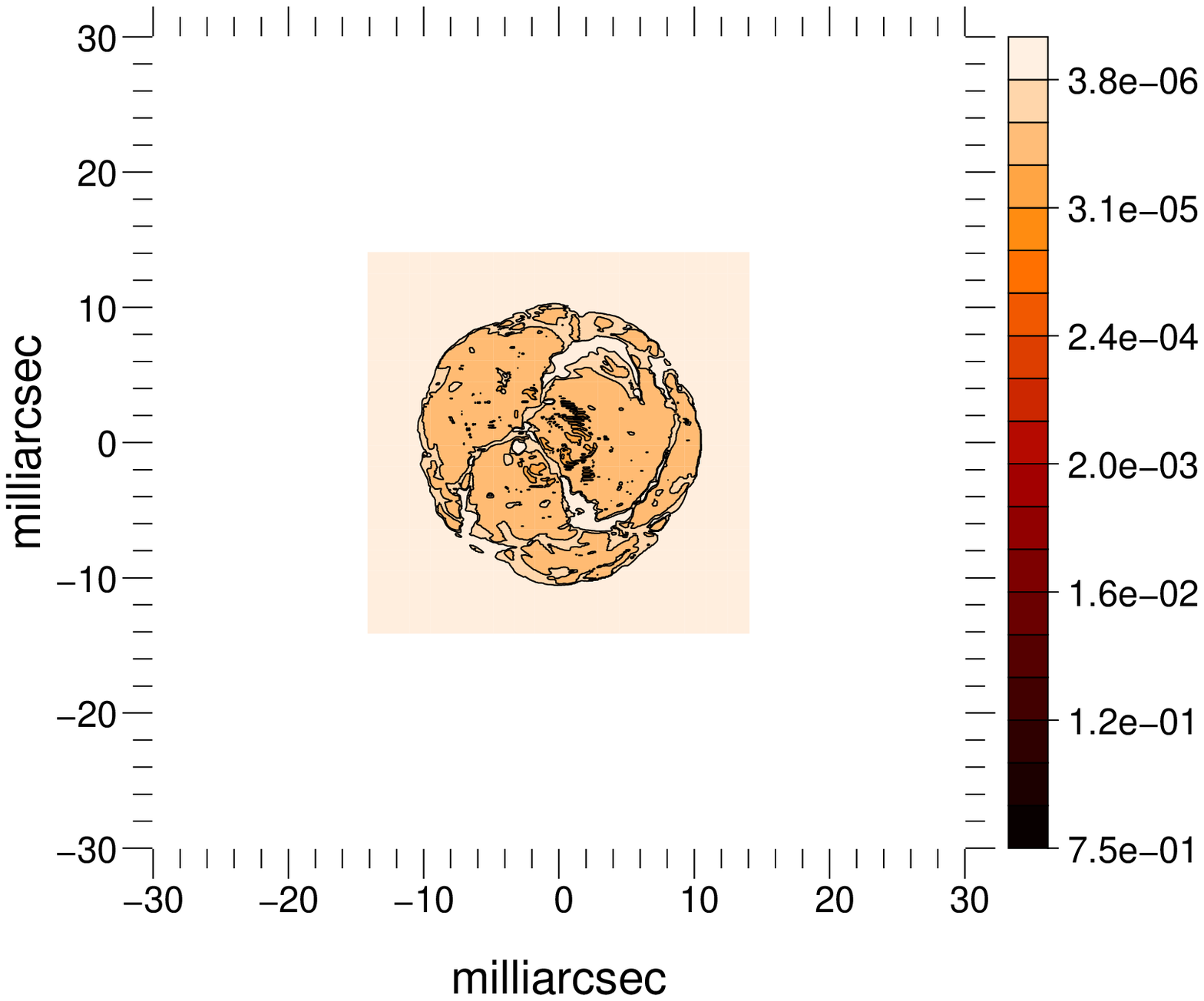}
\includegraphics[height=4cm]{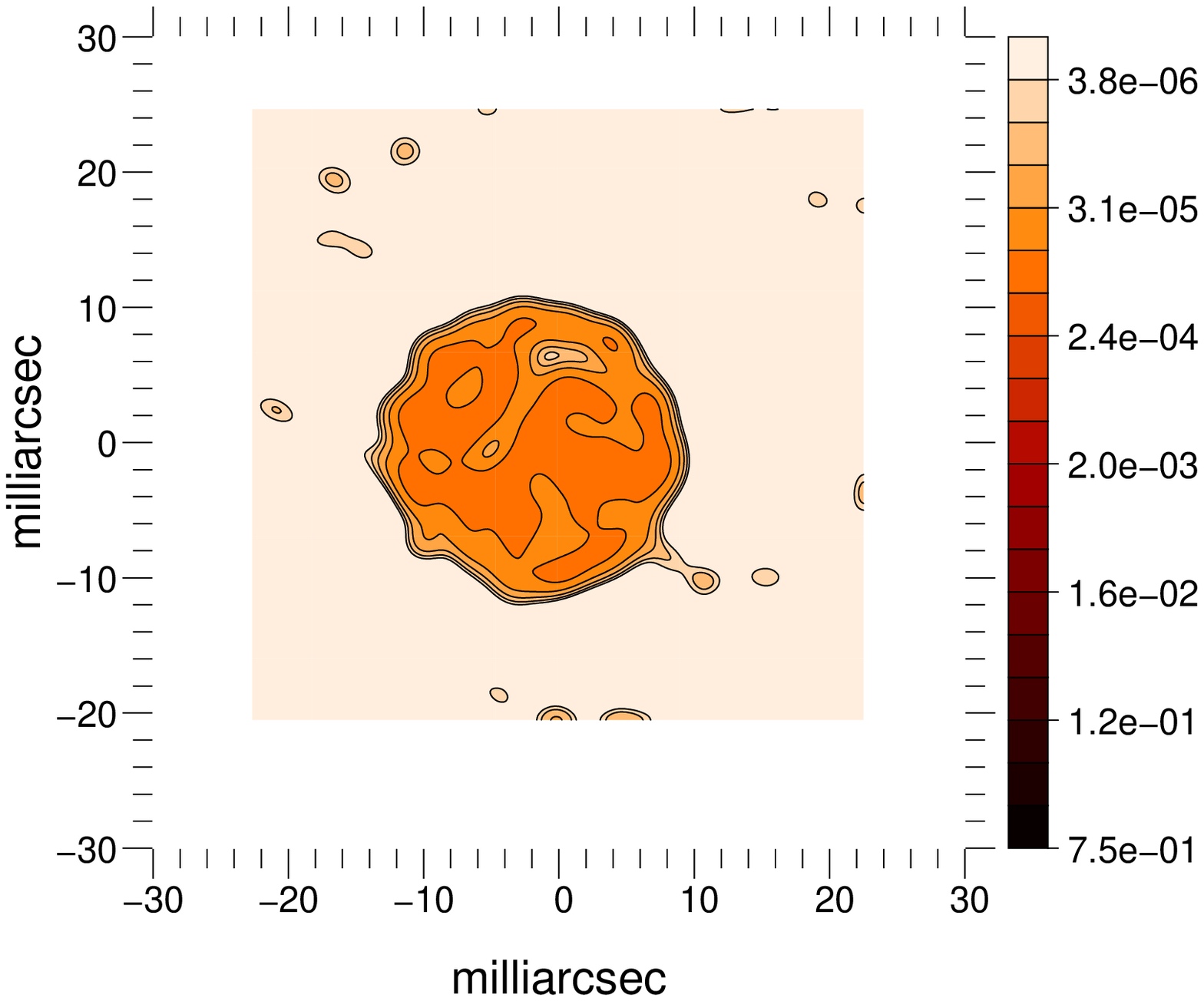}
\includegraphics[height=4cm]{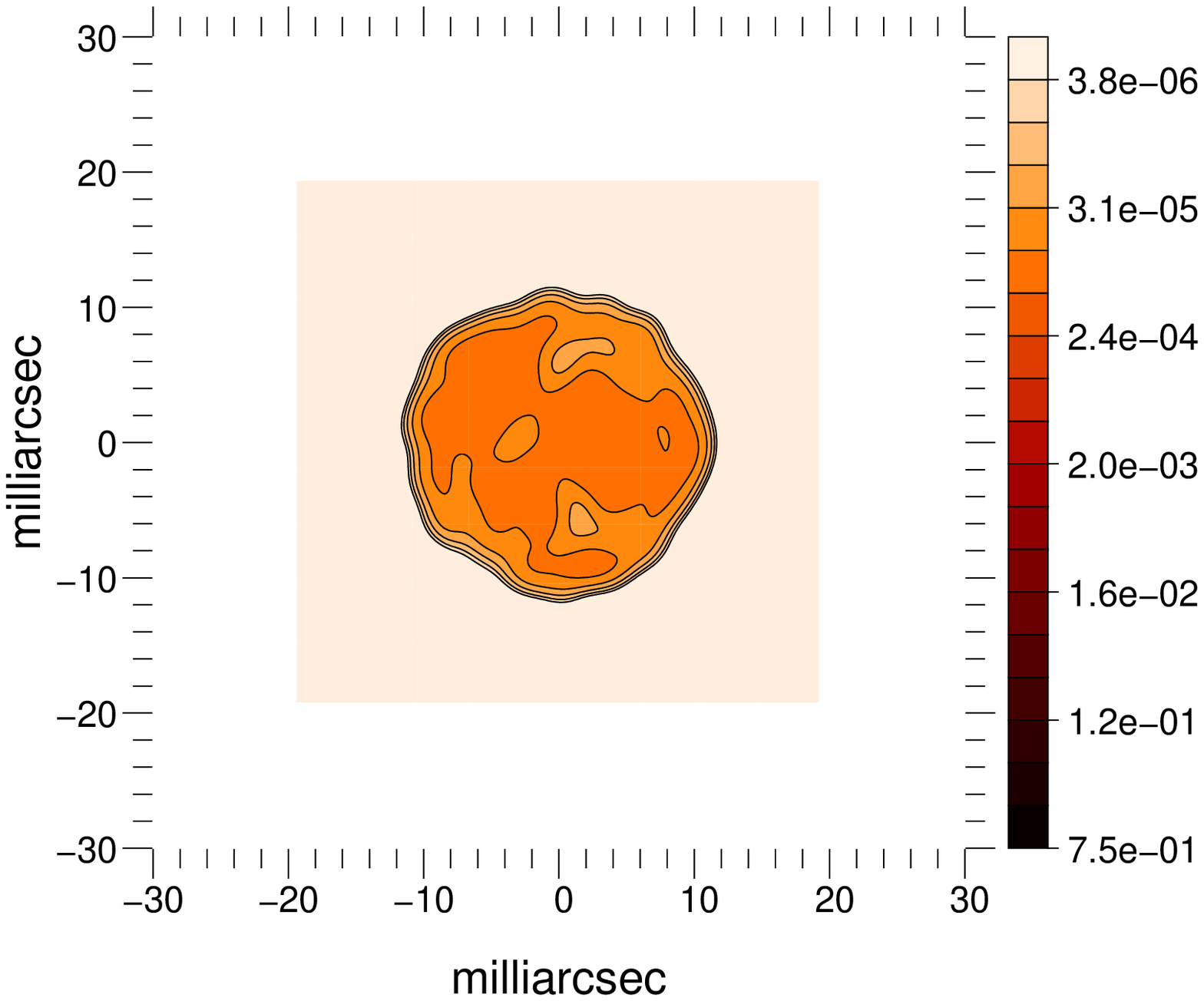}
\end{tabular}
\end{center}
   \caption[SG_surface2] 
   { \label{fig:example} A simulated stellar surface of a supergiant, 0.1 mas/pixel sampling; {\sc mira} reconstruction 4 AT $\times$ 3 nights configuration, smoothness regularization, 0.1 mas/pixel sampling, SNR=$_{V^2}$=4-4900, SNR$_{T}$= 0.004-0.09;  {\sc mira} reconstruction 6 AT $\times$ 1 night configuration, smoothness regularization, 0.1 mas/pixel sampling, SNR$_{V^2}$=4-2999, SNR$_{T}$=0.2-116.}
\end{figure}









\begin{figure}[h!]
   \begin{center}
   \begin{tabular}{c}
 \includegraphics[height=4cm]{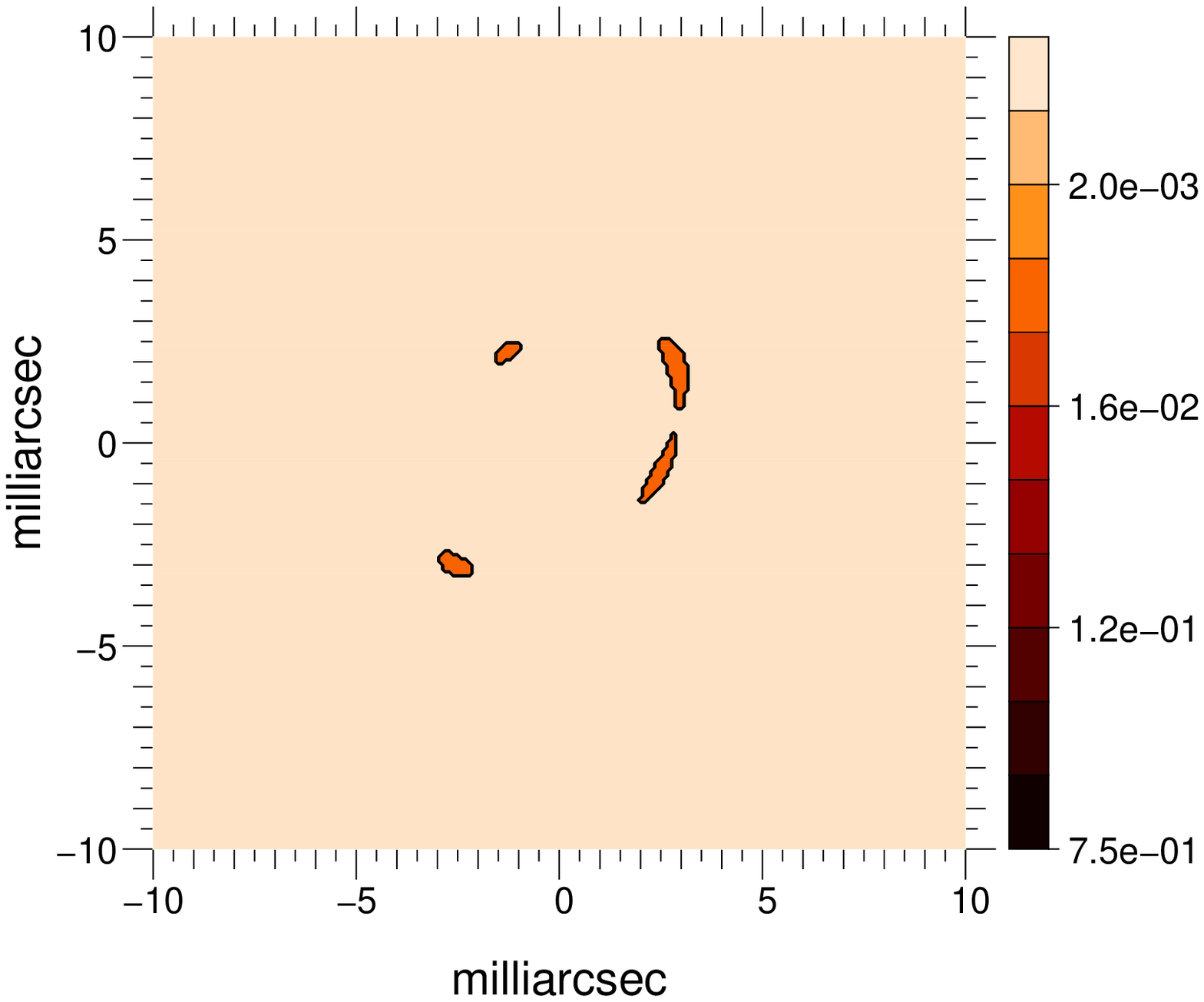}
\includegraphics[height=4cm]{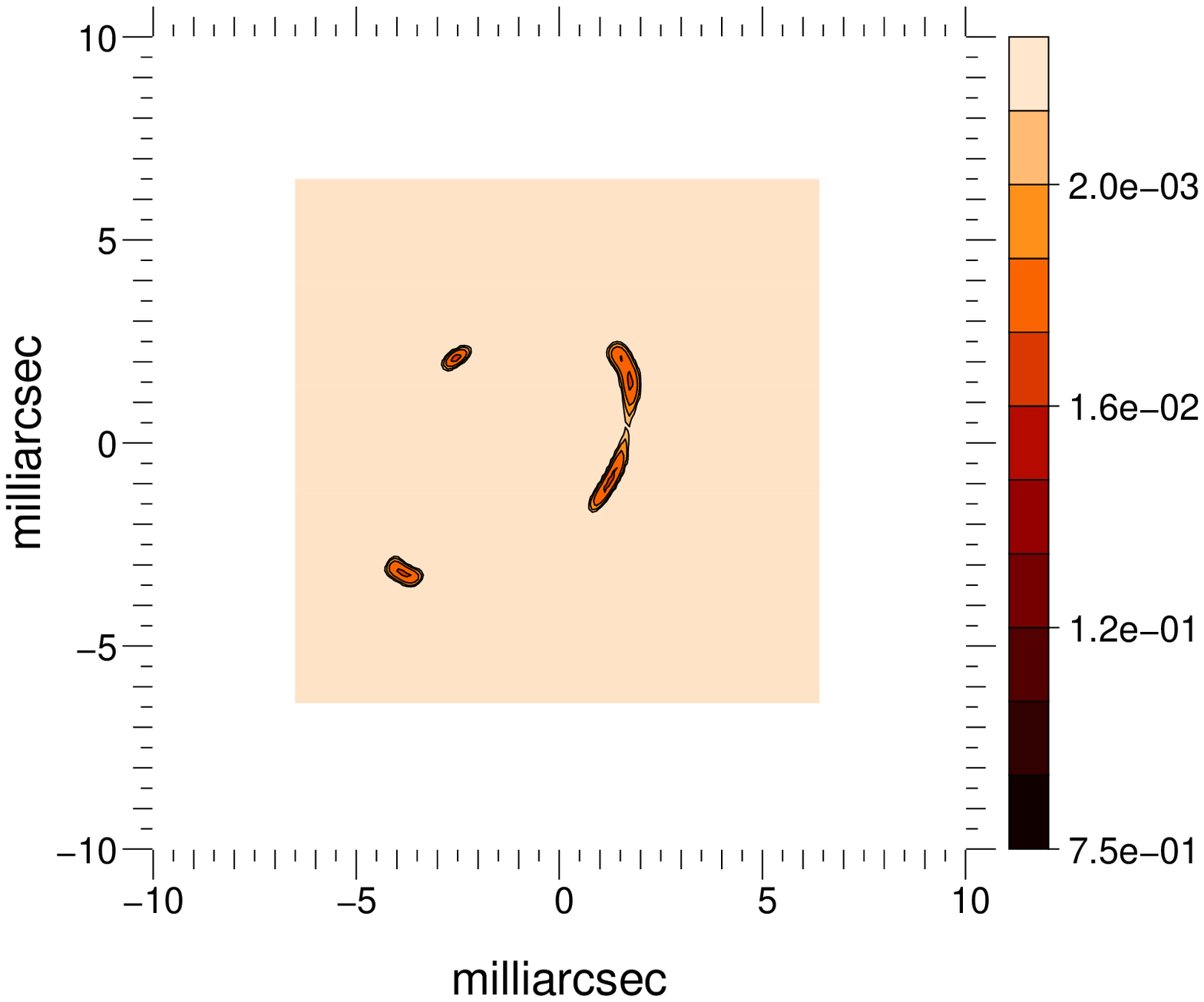}
   \end{tabular}
   \end{center}
   \caption[microlensing] 
   { \label{fig:example} A simulated microlensing event, 0.1 mas/pixel sampling; {\sc mira} reconstruction 4 UT $\times$ 1 night configuration, smoothness regularization, 0.1 mas/pixel sampling.}
\end{figure}

\begin{table}[h!]
\begin{center}
\footnotesize
\caption{Microlensing. Distance is in pixels.}
\begin{minipage}[c]{55mm}
\begin{tabular} {l | c c }

\hline
	&	Image       &	{\sc mira} 4 UT    \\

\hline

A flux	&		19.6\% &	19.6\% \\

B flux	&		11.9\% &	11.9\%	\\

C flux	&		29.8\% &	29.8\%	\\

D flux	&		38.7\% &	38.7\%	\\

ratio D/A &		2.0 &	2.0	\\

ratio D/B &		3.3 &	3.3     \\

ratio D/C &		1.3 &	1.3    \\

\hline

distance AB & 40 & 40 \\

distance BC & 20 & 20 \\

distance CD & 56 & 56 \\

\hline

SNR$_{V^2}$ & - & 228-268 \\

SNR$_{T}$   & - & 0.1-0.4 \\


\hline

\end{tabular}
\end{minipage}
\end{center}
\end{table}





\begin{figure}[h!]
   \begin{center}
   \begin{tabular}{c}
  \includegraphics[height=3.3cm]{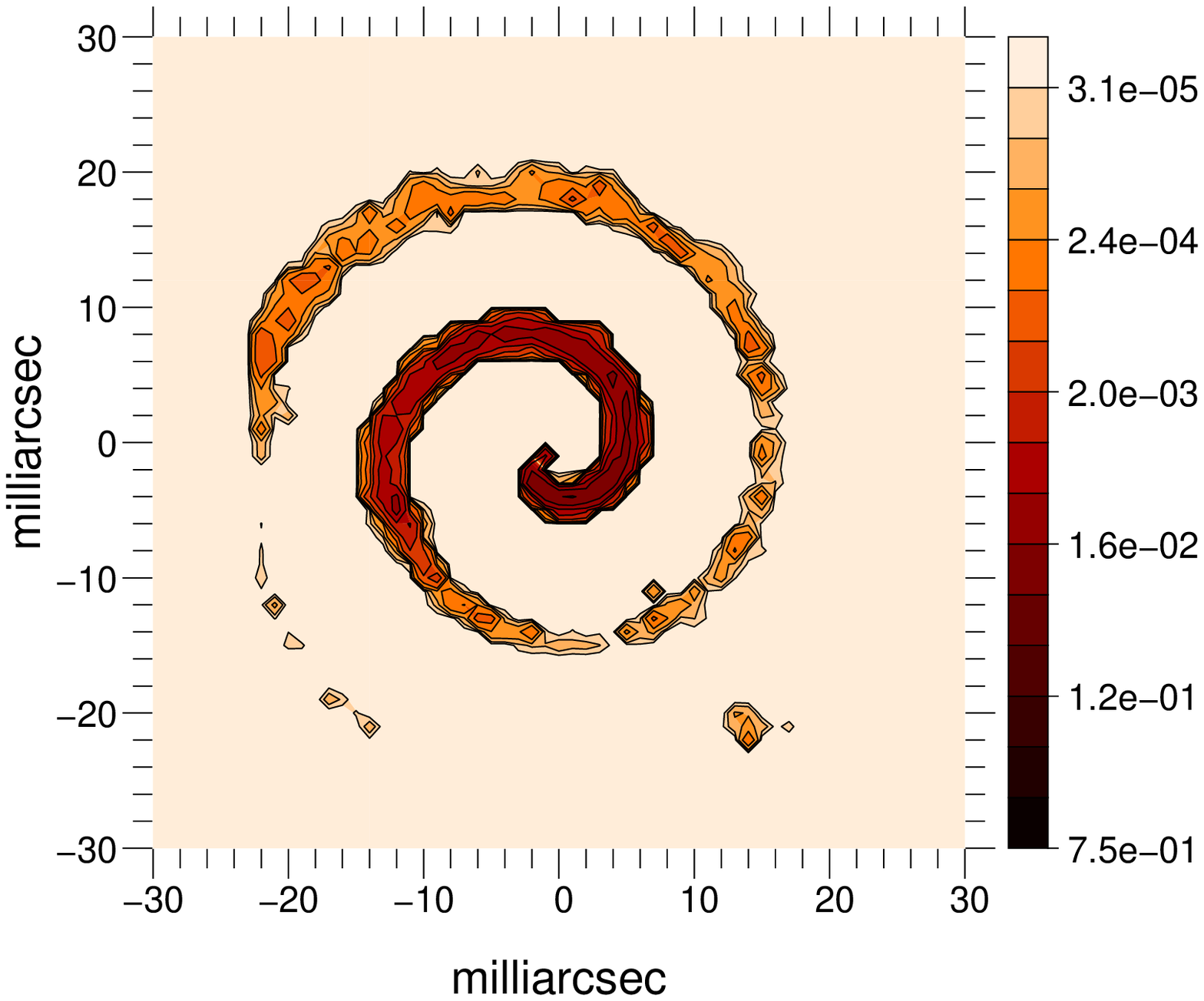}
 \includegraphics[height=3.3cm]{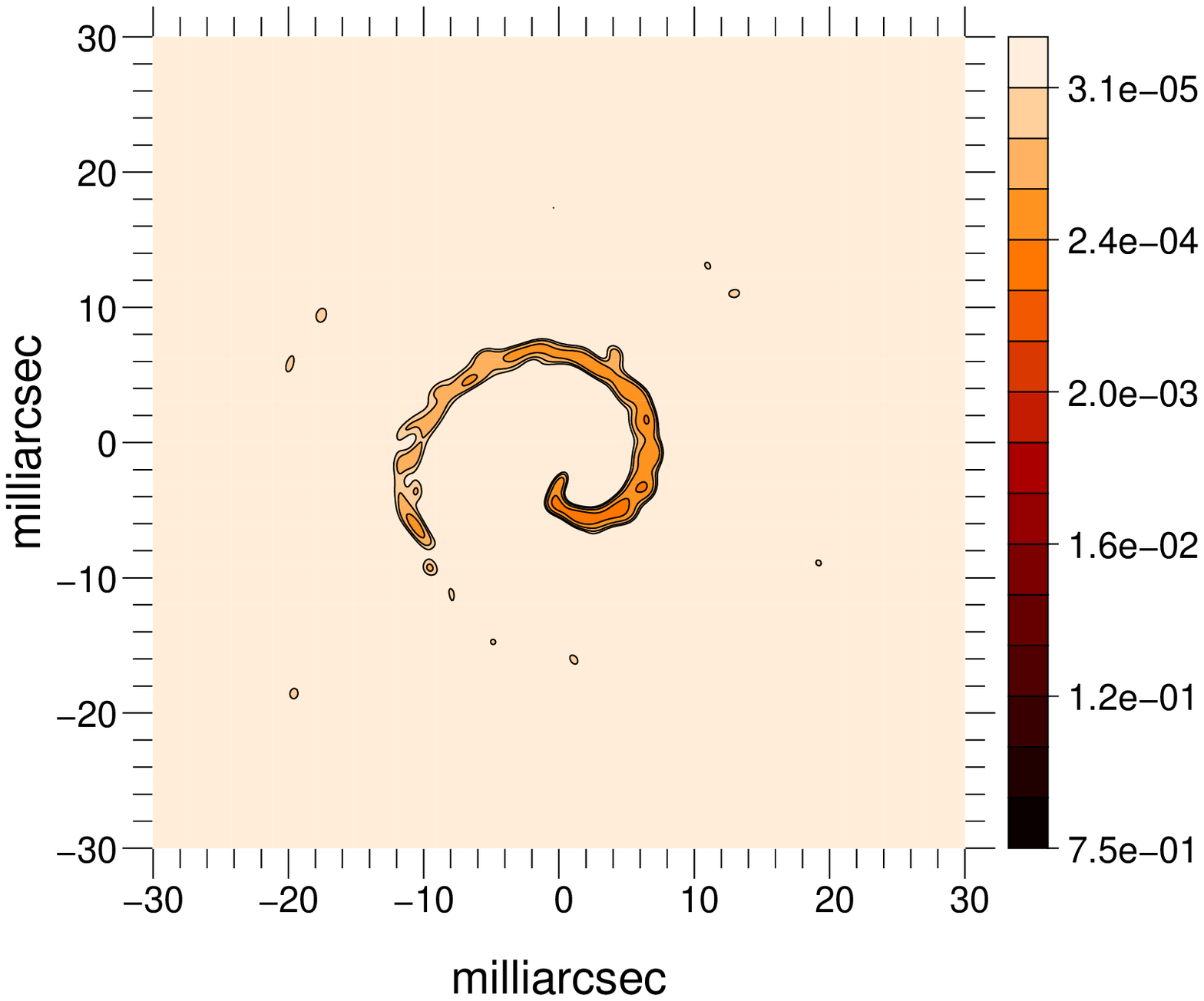}
\includegraphics[height=3.3cm]{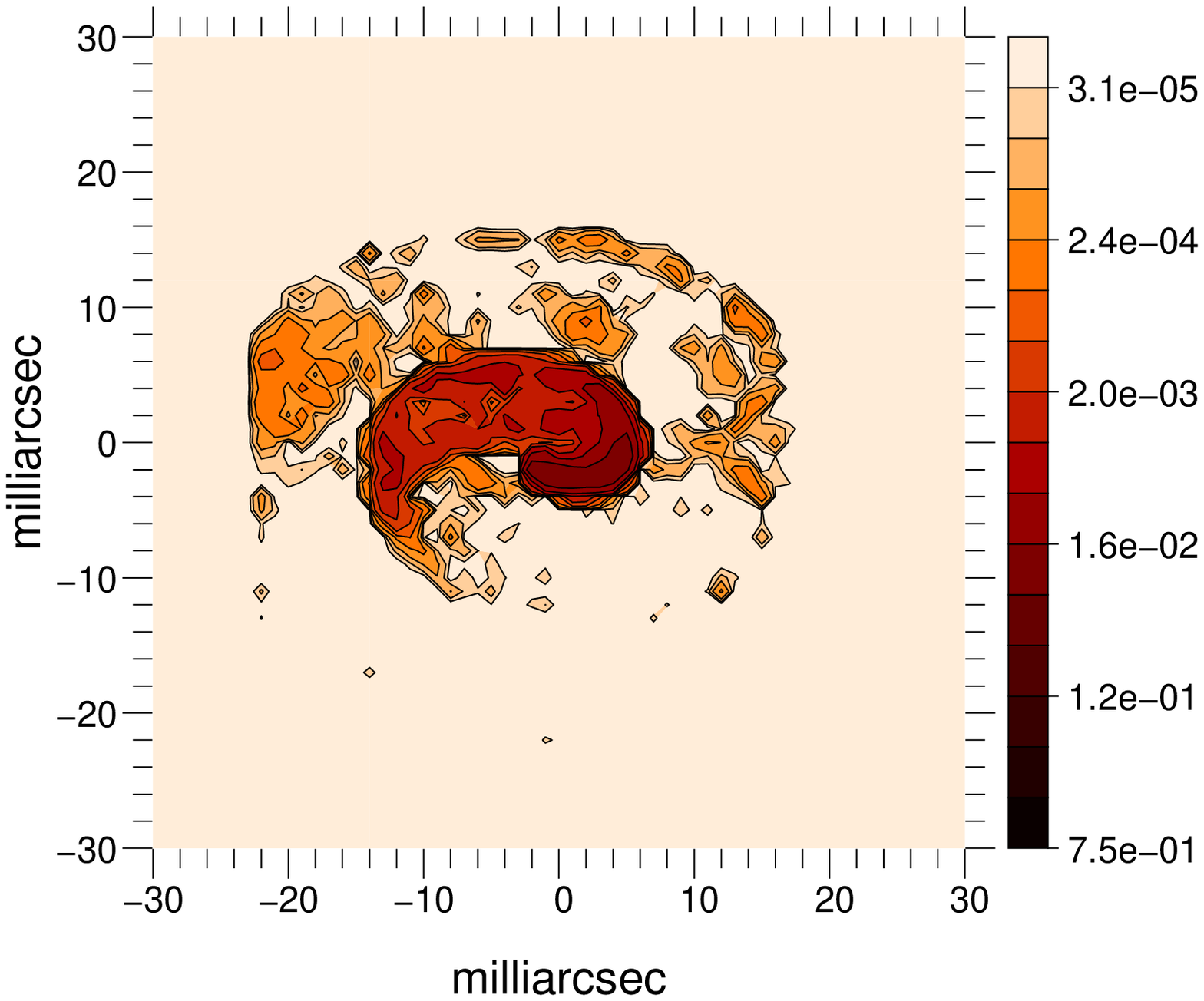}
 \includegraphics[height=3.3cm]{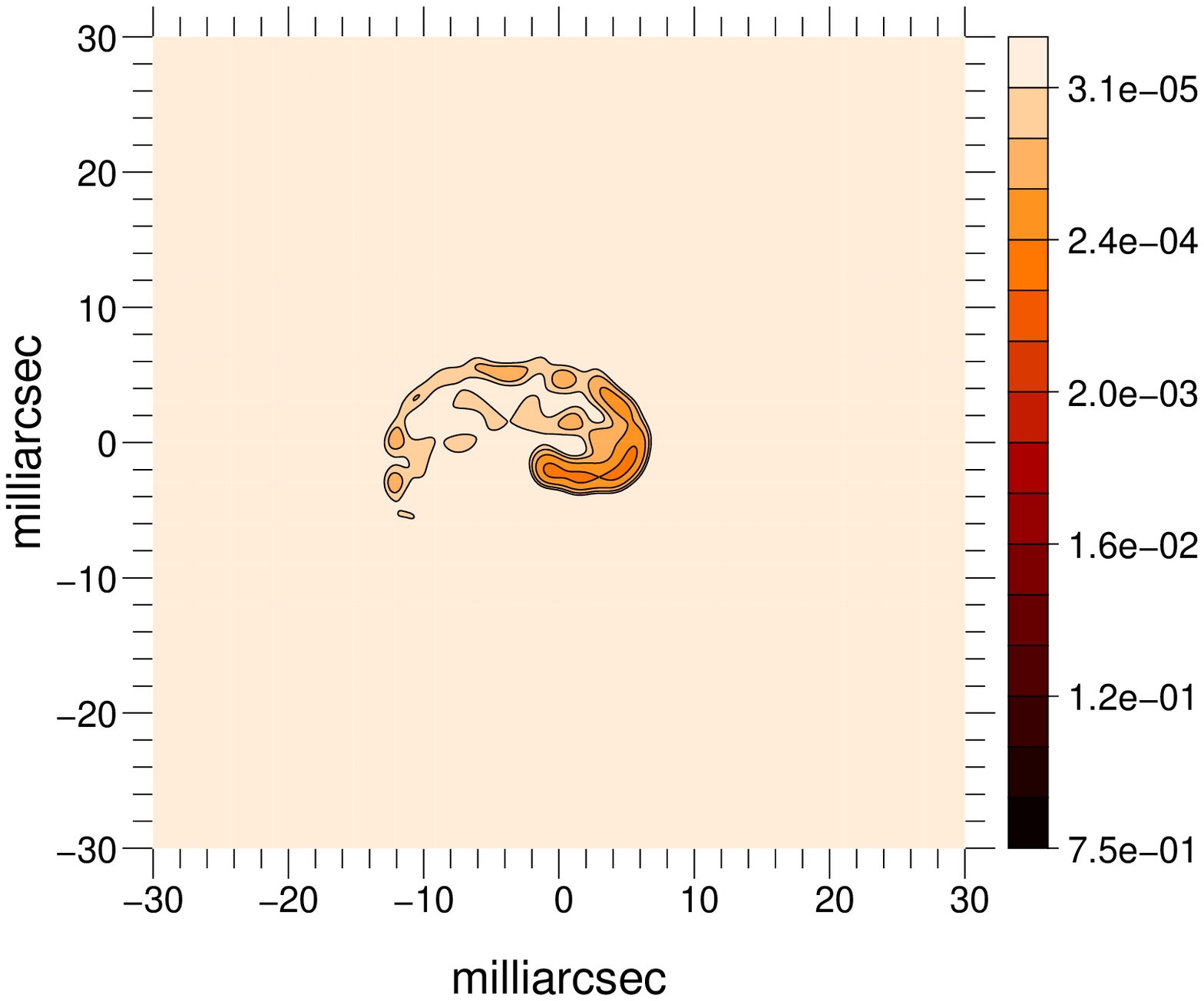}

   \end{tabular}
   \end{center}
   \caption[pinwheel] 
   { \label{fig:example} A simulated image of the pinwheel nebula at 0 degree inclination, 1.0 mas/pixel sampling; {\sc mira} reconstruction 6 AT $\times$ 1 night configuration, smoothness regularization. A simulated iamge of the pinwheel nebula at 60 degree inclination, 1.0 mas/pixel sampling; {\sc mira} reconstruction 6 AT $\times$ 1 night configuration, smoothness regularization, 0.1 mas/pixel sampling.}
\end{figure}

\begin{table}[h!]
\begin{center}
\footnotesize
\caption{Pinwheel}
\begin{minipage}[c]{110mm}
\begin{tabular} {l | c c c c c}

\hline

	& Angle &	Image   & SNR$_{V^2}$ & SNR$_{T}$ & {\sc mira} 6 AT  \\

\hline

inner spiral  & 0 deg &   100 $\times$ 80  & 49-861 & 2-19 & 100 $\times$ 80    \\

inner spiral & 60 deg &   100 $\times$ 60  & 36-861 & 1-43 & 100 $\times$ 60   \\

\hline

\end{tabular}
\end{minipage}
\end{center}
\end{table}

\begin{figure}[h!]
   \begin{center}
   \begin{tabular}{c}
  \includegraphics[height=3.3cm]{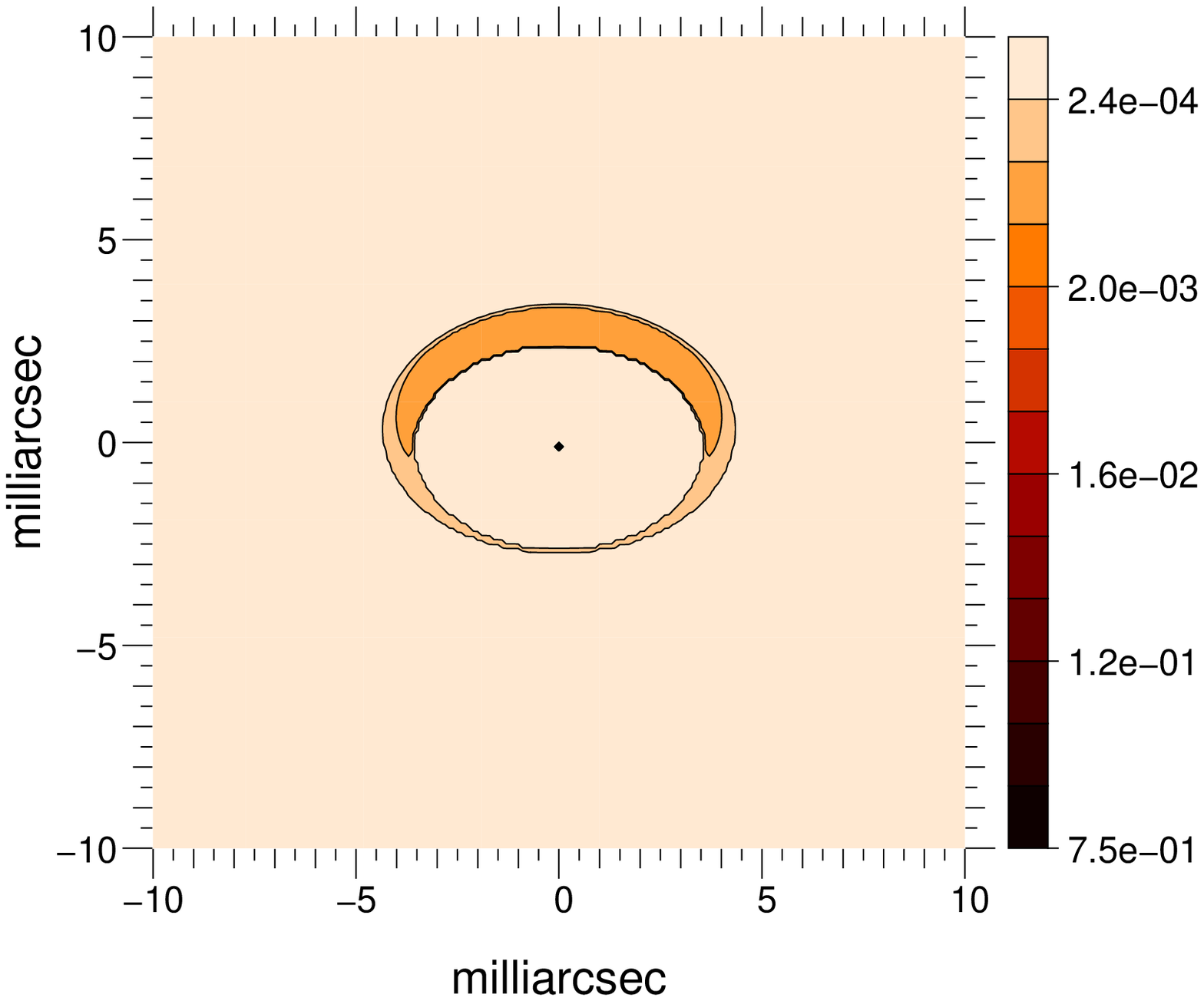}
 \includegraphics[height=3.3cm]{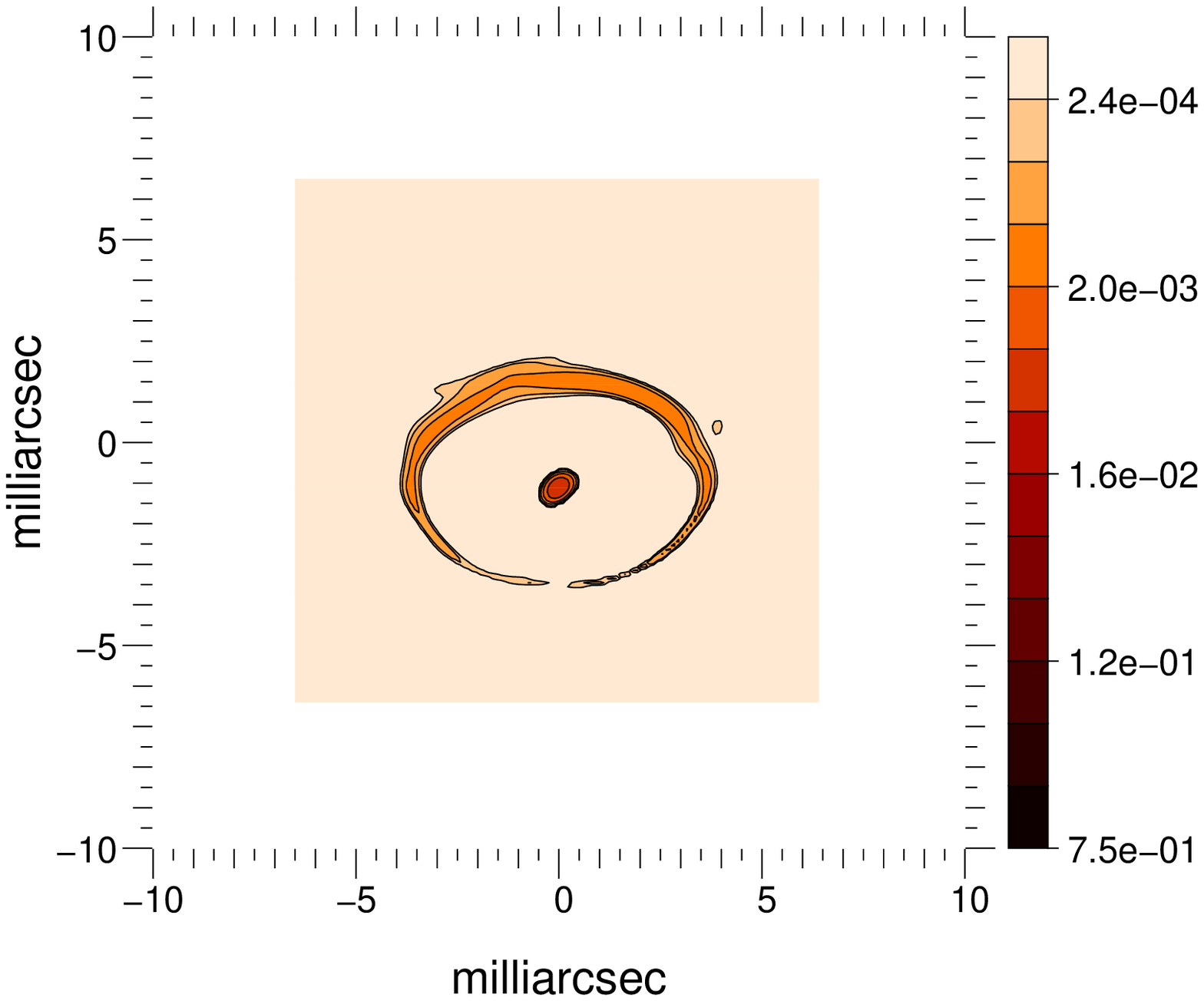}
 \includegraphics[height=3.3cm]{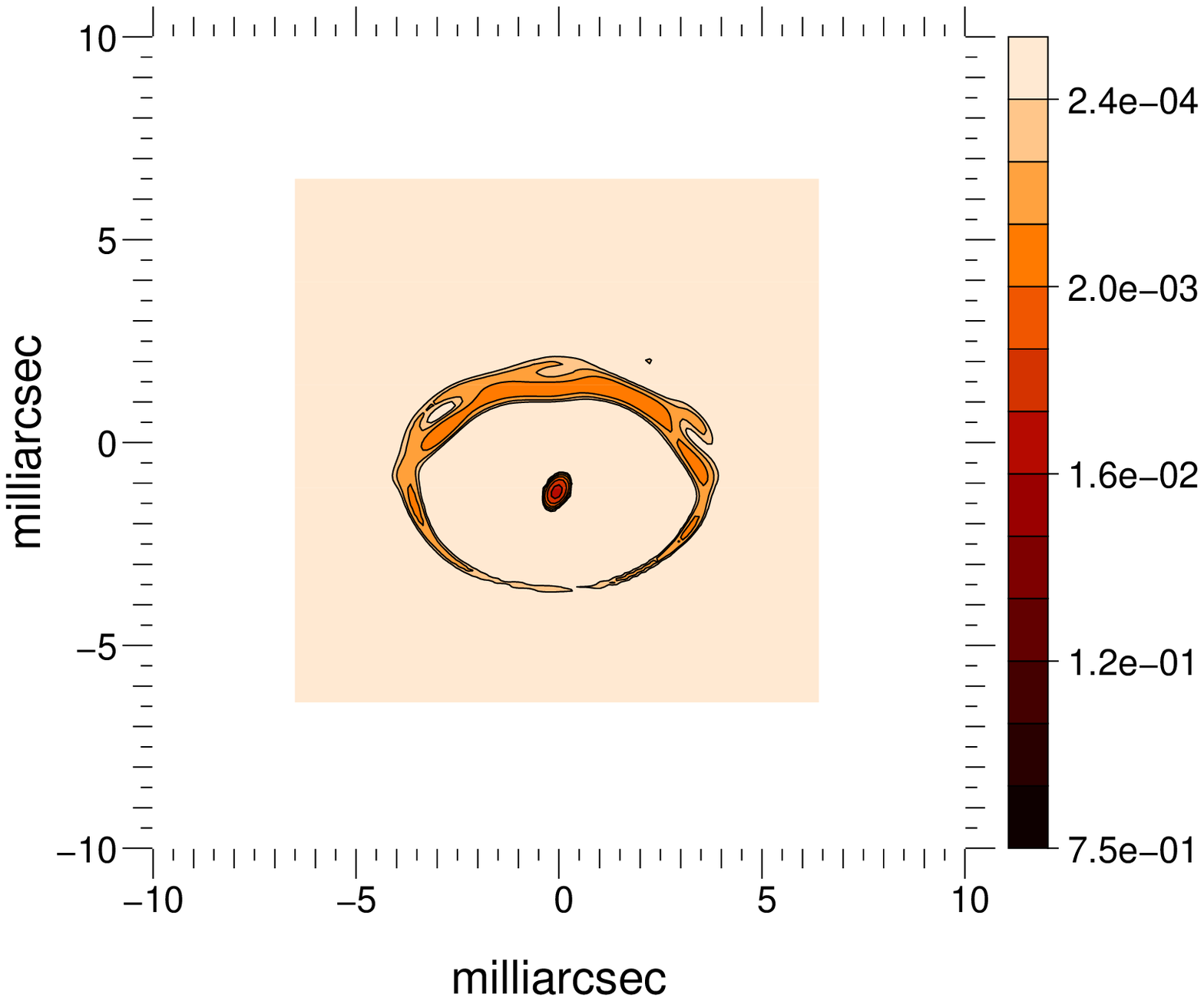}
\includegraphics[height=3.3cm]{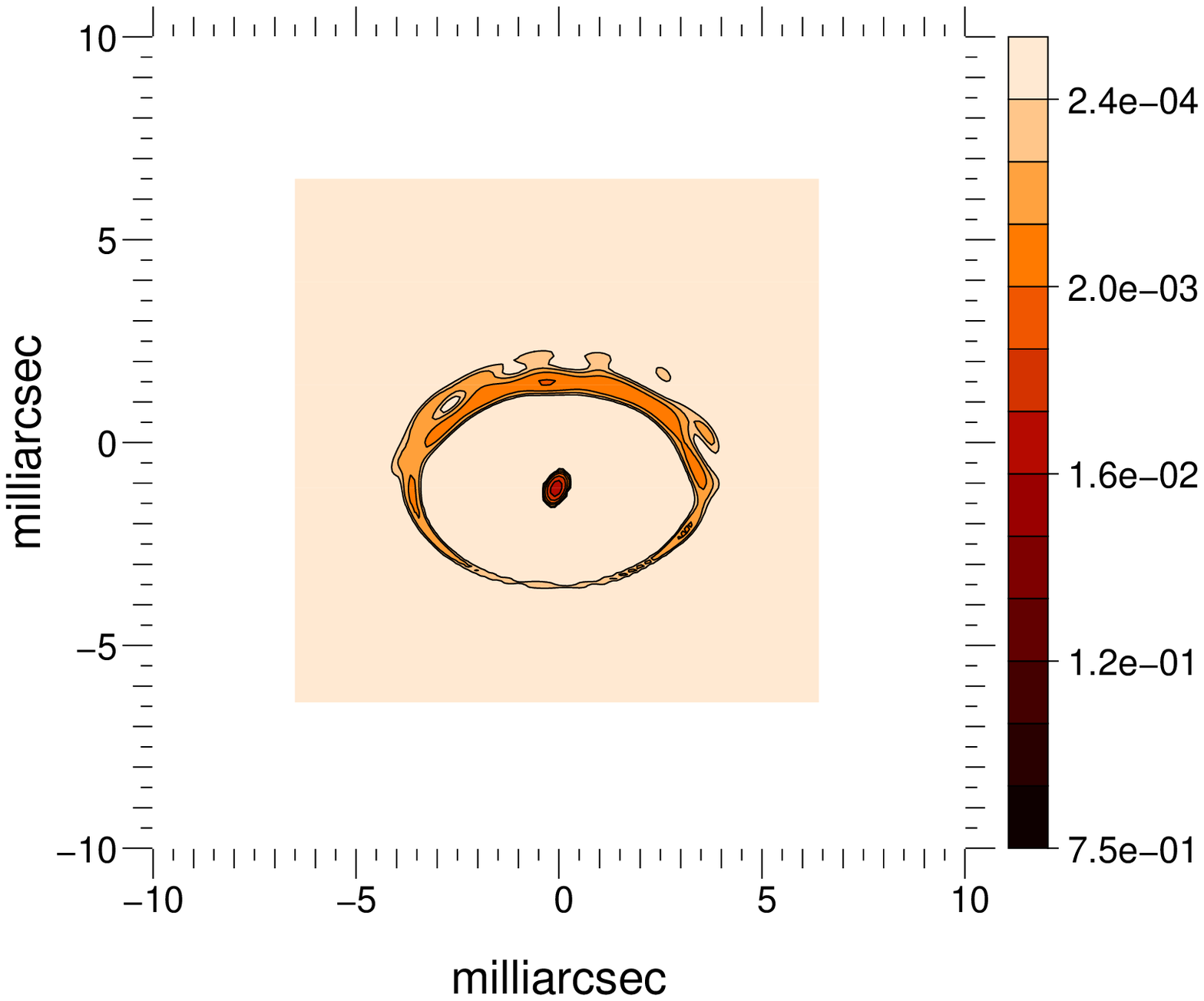}
  \end{tabular}
   \end{center}
   \caption[microlensing] 
   { \label{fig:example}A simulated stellar surface of the structure of inner disks surrounding YSOs, 0.1 mas/pixel sampling; {\sc mira} reconstruction 4 UT $\times$ 1 night configuration, smoothness regularization, 0.1 mas/pixel sampling; {\sc mira} reconstruction 4 AT $\times$ 3 nights configuration, smoothness regularization, 0.1 mas/pixel sampling; {\sc mira} reconstruction 6 AT $\times$ 1 night configuration, smoothness regularization, 0.1 mas/pixel sampling.}
\end{figure}

\begin{table}[h!]
\begin{center}
\footnotesize
\caption{YSO. Radii units are in pixels.}
\begin{minipage}[c]{112mm}
\begin{tabular} {l | c c c c }

\hline
      & Image & {\sc mira} 4 UT & {\sc mira} 4 AT 3 & {\sc mira} 6 AT \\
\hline  
  
flux star & 18.1\%     & 18.7\%	&	18.6\%	&	18.6\% \\

flux disk & 81.9\%     & 81.3\%	&	81.4\%	&	81.4\% \\

ratio     & 0.2        & 0.2	&	0.2	&	0.2 \\

\hline


inner diameter & 65 $\times$ 45 & 65 $\times$ 45 & 65 $\times$ 45 & 65 $\times$ 45 \\    


outer diameter & 85 $\times$ 60 & 80 $\times$ 56 & 82 $\times$ 60 & 85 $\times$ 60 \\

\hline

SNR$_{V^2}$ & - & 2055-5215 & 89-1385 & 12-1363 \\

SNR$_{T}$ &  -  & 0.04-0.2 & 0.04-0.5 & 0.2-12  \\


\hline







\end{tabular}
\end{minipage}
\end{center}
\end{table}







\section {Discussion}







\subsection {The Sources}

In general, the image reconstruction faired very well. This is denoted by the excellent astrometry and photometry of the individual image elements. Some sources merit a separate discussion.

We are able to detect the unresolved AGN out to the sublimation radius. The total flux contained in this region is 91.6\% compared to the 91.4\% measured in the synthetic image. The result suggests that such observations will provide important constraints on the size of the sublimation region and torus in the nearest AGN.

For the stellar surfaces, our results show unprecedented detail. Most of the flux is recovered in a compact region and careful inspection of the reconstructed images shows correspondence with individual regions of different brightness in the synthetic image. The edges of the stellar surface are also well reproduced. 
 
\subsection {Number of Telescopes and $uv$ Coverage}

Our results show that the use of the 6 AT $\times$ 1 night configuration generally yields better reconstructed images than the 4 AT $\times$ 3 nights configuration, even though the number of $uv$ points is similar. This is particularly noted in images of stellar surfaces and in stellar disks/winds. This is in consonance with previous studies (Tuthill \& Monnier 2000; Baron 2007). The image reconstruction result is satisfactory with a low number of array elements when the source is compact (Tuthill \& Monnier 2000). However, when the sources are more complex, a larger Fourier plane coverage is needed and therefore more array elements are needed.

Indeed it is the number of telescopes and not the number of observing nights (and therefore total number of $uv$ points) that determine the image reconstruction outcome. In a system of {\it N} telescopes, the number of {\it uv} points is $\frac{1}{2} N \times (N-1)$ per night per integration point and the number of independent closure phases is {\it $\frac{1}{2}$ $\times$ (N-1) $\times$ (N-2)} per night per integration point. With {\it N=4} and 3 nights, for each integration point there are a total of 18 $uv$ points (therefore squared visibilities) and a total of 9 closure phases. For {\it N=6} and 1 night, the number of squared visibilities and closure phases per integration point is 15 and 10, respectively. 

On the other hand, the image reconstruction (or fitting process) can only make use of the instantaneous phases to create a model brightness. Therefore, for a {\it N=4} system, the image reconstruction uses the 3 instantaneous closure phases (3 per night), while with 6 telescopes there are now 10 independent closure phases.


\subsection {Dynamic Range}

Image reconstruction, whatever the method, is crucial when a high dynamic range ($>$1000) is required. Our results show that the closer phase technique allows to achieve modest dynamic ranges. Higher dynamic ranges can be obtained by using the VLTI coupled with a phase referencing instrument, like PRIMA (Deplanke et al. 2000) which will allow to reach K band magnitudes of 11 mag for the ATs and 14 mag for the UTs. The presence of nearby phase referencing sources will allow larger integration times on source and therefore higher dynamic range.


\section {Conclusions}

Next generation optical interferometric instrumention like the VSI will be designed to offer superb imaging capabilities in order to bypass ambiguities in the interpretation of visibility data. It will be particularly novel in that it will allow spatial resolutions of milliarcseconds, comparable to ALMA and even better the future JWST. 

In order to test future VLTI performance in terms of imaging, we have run several tests on key scientific cases. Results show that with a simple 6 Auxiliary Telescope (AT) configuration and 1 night observation, images of high image fidelity can be obtained. Using the VLTI alone will allow to achieve modest dynamic ranges, although in conjunction with a phase referencing instrument, K band magnitudes of 11 and 14 mag can be reached with the ATs and UTs, respectively. Such  diverse objects as AGN, stellar surfaces and disks are shown possible to image with future VLTI instumentation. 

\section{Acknowledgments}

MEF is supported by the Funda\,c\~ao para a Ci\^encia e a Tecnologia through the research grant
SFRH/BPD/36141/2007.
PJVG and MEF were supported in part by the
Funda\,c\~ao para a Ci\^encia e a Tecnologia through projects
PTDC/CTE-AST/68915/2006 and PTDC/CTE-AST/65971/2006 from POCI, with
funds from the European programme FEDER.

\section {References}

\noindent Baron, F.,  2007, {\sc bsmem} Test Plan Results, private communication


\noindent Deplancke, F. et l. 2000, SPIE, 4006, 365




\noindent Garcia, P. et al. 2007, in doc. VLT-SPE-VSI-15870-4335, issue 1.0 in VSI Phase A Document Package, Science Cases

\noindent Hofmann, K. -H. \& Weigelt, G. 1993, A\&A, 278, 328

\noindent Jocou, L. et al. 2007, in doc. VLT-SPE-VSI-15870-4335, issue 1.0 in VSI Phase A Document Package, System Design

\noindent Lopez, B., et al. 2006, in Advances in Stellar Interferometry, Editted by Monnier, John D., Sch\"oller, Marcus, Danchi, William C., Proceedings of SPIE, 6268, 31
 

\noindent Malbet, F., et al. 2006, in doc. VSI-PRO-001, issue 1.0, VSI Technical Proposal in Response to ESO Call for Phase A Proposals for 2nd Generation VLTI Instruments

\noindent Masoni, L. 2006, PADEU, 17, 155

\noindent Masoni, L. et al. 2005, Astr. Nachr., 326, 566
  
\noindent Pauls, Young, Cotton \& Monnier 2004, Chara Technical Report

\noindent Thiebaut, E., 2005, Optics in Astrophysics: Proceedings of the NATO advanced Study Institute on Optics in Astrophysics, Cargese, France, vol 198, eds. R. Foy \& F. -C. Foy, pg. 397

\noindent Tuthill, P. G. \& Monnier, J. D.,  1 August 2000,  Chara Technical Report, no. 86  

\noindent Weigelt, G., Balega, Y., Bloecker, T., Fleischer, A. J., Osterbart, R., \& Winter, J. M. 1998, A\&A, 333, L51

\end{document}